\documentclass[a4paper,12pt]{article}

\usepackage[a4paper]{geometry}
\usepackage[utf8]{inputenc} 
\usepackage{hyperref}
\usepackage{float}
\usepackage{amssymb}
\usepackage{amsmath}
\usepackage{amsthm}
\usepackage{url}
\usepackage{amsfonts}
\usepackage{graphicx, float, tabularx}
\usepackage{color}
\usepackage{enumerate}
\usepackage[american]{babel}
\usepackage{stackrel}
\usepackage{subfigure}

\usepackage[numbers,sort,compress]{natbib}
\newcommand{\beq}{\begin{equation}}
\newcommand{\eeq}{\end{equation}}

\newcommand{\lagr}{{\cal L}}

\newcommand{\be}{\begin{equation}}
\newcommand{\ee}{\end{equation}}
\newcommand{\bea}{\begin{eqnarray}}
\newcommand{\eea}{\end{eqnarray}}
\newcommand{\mpl}{M_{\rm P}}

\newcommand{\lpl}{L_{\rm P}}
\renewcommand{\d}{\mathrm{d}}

\newcommand{\f}{\frac}
\newcommand{\p}{\partial}

\renewcommand{\frac}[2]{{{\displaystyle #1}\over{\displaystyle #2}}}

\author{Athanasios G. Tzikas \thanks{E-mail: \href{mailto:athanasios.tzikas@unibg.it}{\texttt{athanasios.tzikas@unibg.it}}
} 
\\[1.5ex]
\small \textit{Dipartimento di Ingegneria e Scienze Applicate} \\[-0.5ex]
\small \textit{Universit\`a degli Studi di Bergamo} \\[-0.5ex]
\small \textit{Viale Marconi 5, 24044 Dalmine, Italy} 
}
\date{}

\title{Planckian charged black holes and their cosmological ramifications} 

\begin{document}
\maketitle

\begin{abstract}
The application of nonlinear electrodynamics  at high energy scales has led to a variety of interesting phenomena in recent years, particularly within the context of non-singular spacetime geometries.  Additionally, it is postulated that gravity near the Planck scale is governed by a minimal cut-off length, which acts as a renormalization scale against ultraviolet pathologies. Within this framework, we combine both concepts by introducing modifications to the electric and matter sectors of a black hole  as its size approaches this minimal length. The result is  an electrically charged black hole  that is free from ultraviolet divergences and recovers the Maxwell limit at classical scales. We  further explore the  geometric and thermodynamic properties of the resulting solution within a cosmological anti-de Sitter background, revealing a chemical analogy with that of a Van der Waals fluid. Subsequently, we examine the  charged black hole in de Sitter space and construct four corresponding gravitational instantons.  We then study their cosmological quantum production using the formalism of the pair creation rate within the context of the no-boundary proposal.
\end{abstract}

\thispagestyle{empty}
\newpage

\section{Introduction}
\label{sec:intro}

The concept of nonlinear electrodynamics (NED)  involves the modification of Maxwell's equations at high energy scales, or very short distances, to avoid the appearence of singularities. In 1934 Born and Infeld \cite{BoI34} proposed a fully relativistic NED Lagrangian to cure the divergent electric self-energy of a point charge. A few years later, Heisenberg and Euler \cite{HeE36}, based on Dirac’s vacuum model,  established their own NED theory, incorporating one-loop quantum corrections to quantum electrodynamics. In 1969, a further extension of the Born-Infeld model was proposed by  Pellicer and Torrence \cite{PeT69}.  

In recent years, there has been considerable interest in models combining NED with gravity. For instance, magnetically charged black hole solutions inspired by NED Lagrangian functions \cite{Bre05b,Kru17,Kru18}, or regular (non-singular) black holes and solitonic ones obtained from the so-called FP-duality \cite{Bro01,Bro22}. Other interesting concepts involve NED-inspired wormholes \cite{Bro17}, dyonic configurations \cite{Bro17b,Pan20}, NED colliding waves \cite{GaB96,Bre96} and electrically charged non-linear configurations satisfying continuous probability distributions \cite{BaV14b}, as well as,  the weak energy condition \cite{BaV14}.  Moreover,  NED can  improve the initial Big Bang singularity providing non-singular FRW cosmological models \cite{GaB00,CGC++04,NGS++07}, as well as,  NED-inspired models of an inflating universe \cite{Kru17c,Kru20}. 
It is worth mentioning that Beato and Garcia \cite{AyG00} derived via NED the Bardeen black hole, which is the first non-singular black hole solution  proposed in the literature \cite{Bar86}. Their derivation resulted from the strong coupling of  a magnetic monopole with the graviton.  Apart from this, they introduced regular black holes that are coupled to both an electric  and a magnetic monopole and obey  the Maxwell theory for weak fields \cite{AyG99a,AyG99b,AyG99c}.

On the other hand, the success of general relativity is beyond questioning, especially in light of recent experimental observations. These include gravitational wave detection by LIGO and Virgo \cite{GW15,GW17,GW19} and the capture of the first image of a supermassive black hole by the Event Horizon Telescope \cite{EHT19}. Nevertheless, the theory lacks a full quantum description due to its non-renormalizable character near the Planck scale \cite{Sho07}. This limitation becomes apparent at short distances, such as the singular center of a black hole or the initial Big Bang singularity. A complete   theory of quantum gravity is anticipated to resolve these issues, including the cosmological constant problem, the hierarchy problem and the trans-Planckian problem \cite{Rov04}. String theory \cite{GSW87} and loop quantum gravity \cite{Rov98} are two potential candidates but both lack  experimental proof. Other notable attempts involving higher and lower dimensional scenarios are under consideration  \cite{ADD98,AAD+98,tHo93,TNMC18}. 

In the context of black hole physics, effective approaches are of particular interest \cite{Mea64,Dym92,Dym03,BoR00,Mod05,Hay06,BDM07,Ans08,NiW11,MMN11,LeZ11,IMN13,SpS15,CMN15+,SpS17,SpS17D,Nic18},  as they incorporate the feature of  a minimal cut-off length $\vartheta$ that acts as a renormalization scale against ultraviolet divergences. 
The emergence of $\vartheta$ originates from quantum fluctuations anticipated near the Planck scale, which are encoded in the uncertainty of spacetime itself. The spacetime uncertainty, in turn, inherits the feature of non-locality in our measurements \cite{EMK+91,MoR17}. As a result, quantum fluctuations forbid experimental observations below this length and prevent the source to collapse into a singular matter point.  This implies the existence of a central repulsive core that replaces the curvature singularity of a black hole and acts against gravitational collapse. 

Among the spacetimes inspired by the existence of $\vartheta\,$, a notable example is the family of noncommutative black holes \cite{Nic09}. These objects incorporate  short-distance quantum effects to the matter sector by introducing a Gaussian narrow distribution around the center, rather than a Dirac-delta mass function. This way, the mass can be visualized as a spherical, self-sustained, anisotropic fluid of radius $\vartheta$ that is diffused throughout the origin. Consequently, point-structures lose their meaning and are replaced by smooth distributions. Noncommutativity has provided ultraviolet improved manifolds, free from curvature singularities, that are neutral \cite{NSS06}, charged \cite{ANS++07}, rotating \cite{MoN+10}, lower \cite{MuN11,GNS20} and higher dimensional \cite{SSN09}. 

Similar phenomenology with the smeared-like anysotropic matter fluid can be attained through other approaches, such as generalized uncertainty principle  effects  \cite{Nic12,CMN15,KKM+19}, string T-duality \cite{NSW19,GaN22,GKN22} and holographic screens \cite{NiS14}.  All these black hole spacetimes possess an inner repulsive core while recovering general relativity at low energies. Furthermore,   other  intriguing models involve purely quantum mechanical description of microscopic black holes, provided they respect the classical limit of general relativity. Some of these models are governed by the idea of the quantum N-portrait \cite{DvG13,FKN16}, the horizon wavefunction formalism \cite{Cas13,CGO16} and the dynamical particle-like description of microscopic black holes \cite{SpS16++,SpS16}.

In the spirit of the above description, we make the ansatz of  a static  electrically  charged matter source, such as an anisotropic fluid, located at the origin of spherical coordinates. The source, generating a charged black hole, needs to ensure regularity for both  the electric  and the matter sector without necessarily invoking strong coupling between the electric monopole and the graviton. This distinguishes our approach from the papers on NED mentioned earlier.  For this reason, we conjecture that short-distance quantum fluctuations  affect not only gravity but also electrodynamics in a distinctive manner. More precisely, we will consider two separate Lagrangians (one for the matter sector and one for the electric sector) rather than a single Lagrangian that involves the strong coupling of the source’s bare mass and charge. To put it another way, it is the existence of the minimal length $\vartheta$  that triggers the onset of NED corrections and, at the same time, removes  the curvature singularity associated with the matter sector. The value of $\vartheta$ can be arbitrary, \textit{e.g.} the Planck length $\lpl$ or the minimal length of a string, but for our purposes we will choose $\vartheta = \lpl \sim 10^{-33} \mathrm{cm}$. 
Being that said, our charged matter source can be visualized  as a self-sustained entity  where  quantum fluctuations  prevent not only the matter field but also the electric field from diverging at the center.  Along this line of reasoning, spacetime below the Planck length is essentially  meaningless, as it cannot be tested experimentally, and thus, ultraviolet divergences are effectively avoided. Therefore, the space is divided into two distinct regions: a physically meaningful region for $r \geq \lpl$ and a physically meaningless region for $r < \lpl$. As we delve into the subject, intriguing phenomena become apparent regarding the behaviour of these objects within a cosmological background.

The structure of the paper is as follows: in Sec.~\ref{sec:rbh}, we begin with the gravitational action that describes our system. From its variation, we solve the resulting field equations by specifying the form of the NED and the matter Lagrangian. The solution yields an ultraviolet improved black hole governed by NED effects in the quantum regime, while  retrieving the conventional Reissner-N\"orsdtrom metric in the classical regime. Apart from this, we examine the validity of the energy conditions, the thermodynamic properties of the black hole and  the possible onset of  Schwinger pair production close to the event  horizon. 
In Sec.~\ref{sec:ads_rbh} and Sec.~\ref{sec:ds_rbh}, we place the black hole inside a cosmological background governed by a cosmological constant $\Lambda\,$.  The sign of $\Lambda$ can be either negative or positive, indicating the topology of  anti-de Sitter (AdS) or  de Sitter space, respectively. For the AdS configuration (Sec.~\ref{sec:ads_rbh}), we find a chemical analogy between the black hole and the Van der Waals gas. This analogy becomes possible within the framework of black hole chemistry \cite{KuM14}, where $\Lambda$ is interpreted as  the thermodynamic pressure of the system. 
For the de Sitter configuration (Sec.~\ref{sec:ds_rbh}), we find four different solutions (lukewarm, Nariai, cold and ultracold black hole) that exhibit topological horizon similarities with their classical  counterparts \cite{Rom92}. In Sec.~\ref{sec:pbs}, we employ the concept of the no-boundary proposal \cite{HaH83} to investigate the quantum production  of each of the aforementioned   non-singular de Sitter solutions. Based on the analysis of Refs.~\cite{BoH95,MaR95,BoH96}, we formulate the expression of the pair creation rate $\Gamma$ and investigate possible black hole pair\footnote{For more details on the mathematical structure of the nucleated black hole pair see \cite{GKP06}.} nucleation during and after inflation.   
In Sec.~\ref{sec:concl}, we summarize our results and give insights for future work. Throughout the paper we use natural units by setting $c=\hbar=k_{\rm{B}}=1$ and  $\mu_0=4\pi$ so that $G=\lpl^2=\mpl^{-2}$.

\section{Asymptotically flat solution}
\label{sec:rbh}

The concept of NED implies  short scale modifications of Maxwell's theory by replacing  the scalar field strength $F_{\rm M}$ in the conventional Einstein-Maxwell action with a suitable Lagrangian function of it, i.e., $\lagr = \mathcal{L}(F_{\rm M})$. The subscript ``M" stands for ``Maxwell''. This Lagrangian should satisfy two limits:
\begin{enumerate}
\item The weak field limit: For weak fields, or large distances,  the conventional Maxwell theory must be recovered $\left( \lagr \approx F_{\rm M}   \right)$.
\item The regularity limit: For strong fields, or short distances,  a regular core and, consequently, a regular Lagrangian must be present, avoiding this way ultraviolet divergences ($\mathcal{L} \approx \rm const.$).
\end{enumerate}
The starting point of our analysis is the general action
\begin{equation} \label{action}
\mathcal{S} =  \int \d ^4 x \sqrt{-g} \left[  \frac{ R-2\Lambda }{16 \pi \lpl ^2}  - \frac{\mathcal{L}}{4\pi}  +  \mathcal{L}_{\rm m} \right] ,
\end{equation}
where $R$ is the Ricci scalar, $\lagr$ is the the NED Lagrangian and $\lagr_{\rm m}$ is the matter Lagrangian. These forms will be specified below, after providing the necessary elements for their determination. It is noteworthy that, in contrast to other actions inspired by NED, which typically feature a single Lagrangian containing both the mass and the monopole charge, our formulation involves two separate Lagrangians; one includes the mass  of the source, while the other  the electric monopole  independently from the mass. The variation of \eqref{action} with respect to the electromagnetic four-potential $A_{\mu}$ and the metric tensor $g_{\mu\nu}$ yields two field equations; a Maxwell-like and the Einstein field equation reading respectively
\begin{equation}\label{field_eqs}
\f{1}{\sqrt{-g}} \p _{\mu} \left( \sqrt{-g} \lagr_F F^{\mu\nu} \right) =  0  \qquad \mathrm{and} \qquad  R^{\mu}_{\nu} - \f{R}{2} \delta^{\mu}_{\nu} + \Lambda \delta^{\mu}_{\nu} = 8\pi \lpl^2 T^{\mu}_{\nu} \,
\end{equation}
  with  $\lagr_F = \p \lagr / \p F\,$. For this section we set $\Lambda=0\,$.
The field strength $F_{\mu\nu}$ in a curved, torsion-free spacetime  with the ($-+++$) signature convention takes the  form of
 \begin{equation} \label{F_str}
 F_{\mu\nu} = \p_{\mu} A_{\nu} - \p_{\nu} A_{\mu} = \left( \delta_{\mu}^t \delta_{\nu}^r -  \delta_{\mu}^r \delta_{\nu}^t   \right) E(r) \,,
 \end{equation}
where 	we consider  $A_{\mu}=(A_{t},0,0,0)\,$. The component $A_{t}$ represents the scalar potential. The trace of \eqref{F_str} is connected with the radial  electric field $E(r)$ through the relation
\begin{equation}\label{F}
F = \frac{1}{4} F_{\mu\nu} F^{\mu\nu} = - E^2(r)/2  \,.
\end{equation}
In the case of electric monopoles, the trace $F$ is subjected to NED corrections itself and differs from the Maxwell field strength $F_{\rm M}\,$ ($F \neq F_{\rm M}$). Of cource, in the weak field limit, the condition $\lagr \approx F \approx F_{\rm M}$ must be satisfied. This occurs because the  electric field function $E(r)$  is connected to the $tr-$component of $F_{\mu\nu}\,$.  Conversely, in the case  of magnetic monopoles, the  magnetic field function $B(\theta)$ resides in the $\theta\phi-$component of $F_{\mu\nu}$, and thus, no radial modification of $F$ is needed. This fact is a consequence of the \textit{Dirac monopole}\footnote{According to Dirac \cite{Dir31}, the magnetic monopole is not considered  an independent self-sustained particle. Instead, it lies at the end-point of an infinitely long solenoid (Dirac string). This configuration ensures the validity of the Maxwell equation $\vec{\nabla} \vec{B}_{\rm total}=0$, preserving the closed nature of the total magnetic field ($\vec{B}_{\rm total} = \vec{B}_{\rm monopole}+ \vec{B}_{\rm string}\,$).}. 

The stress-energy tensor $T^{\mu}_{\nu}$ in \eqref{field_eqs} consists of two parts: a matter tensor $T^{\mu}_{\nu}|_{\rm m}$ and an electric tensor $T^{\mu}_{\nu}|_{\rm e}$ reading 
\begin{equation} \label{stress_tensors1}
T^{\mu}_{\nu}|_{\rm m} = \f{-2}{\sqrt{-g}} \f{\delta \left( \sqrt{-g} \lagr_{\rm m} \right) }{\delta \delta^{\mu}_{\nu}} \qquad \mathrm{and} \qquad  T^{\mu}_{\nu} |_{\rm e} = \f{1}{4\pi} \left( \lagr_F F^{\mu \lambda} F_{\nu \lambda} - \lagr \delta^{\mu}_{\nu} \right) 
\end{equation}
such that $T^{\mu}_{\nu} = T^{\mu}_{\nu}|_{\rm m} + T^{\mu}_{\nu}|_{\rm e}\,$. Both tensors are diagonal matrices obeying the anisotropic forms 
\begin{equation} \label{stress_tensors2}
T^{\mu}_{\nu}|_{\rm m} = \mathrm{diag} \left( -\rho_{\rm m}(r),p_r^{(\rm m)}(r), p_{\bot }^{(\rm m)}(r), p_{\bot }^{(\rm m)}(r) \right) 
\end{equation}
and
\begin{equation} \label{stress_tensors2b}
 T^{\mu}_{\nu}|_{\rm e} = \mathrm{diag} \left( -\rho_{\rm e}(r),p_r^{(\rm e)}(r), p_{\bot }^{(\rm e)}(r), p_{\bot }^{(\rm e)}(r) \right) .
\end{equation}
The anisotropy  of the charged fluid-source is necessary because it introduces two pressure terms;  one tangential pressure $p_{\bot}(r)$ and one radial pressure $p_r(r)$ that balances the gravitational attraction, preventing the source from collapsing into a singular matter point.  The divergence-free condition for each tensor separately, i.e., $\nabla_{\mu}T^{\mu}_{\nu}|_{\rm m}=0$ and $\nabla_{\mu}T^{\mu}_{\nu}|_{\rm e}=0\,$, implies the energy-momentum conservation, along with the following conditions:
\begin{equation} \label{pressures}
 p_r^{(\rm m),(\rm e)}(r) = - \rho_{\rm m,e}(r)  \qquad \mathrm{and} \qquad   p_{\bot }^{(\rm m),(\rm e)}(r) = -  \rho_{\rm m,e}(r) - \f{r}{2} \p_r \rho_{\rm m,e}(r) \,.
\end{equation}
Imposing now a static and spherically symmetric line element of the form
\begin{equation} \label{line_el}
\d s^2 = - f(r) \d t^2 +  \f{\d r^2}{f(r)} + r^2 \left( \d \theta ^2 + \sin^2\theta \ \d \phi^2 \right)  
\end{equation}
with a metric potential
\begin{equation} \label{metric}
f(r) = 1 - \f{2m \lpl^2}{r} h(r) + \f{Q^2 \lpl^2}{r^2} g(r) \,,
\end{equation}
we can solve the  field equations \eqref{field_eqs}. The  dimensionless  functions $h(r)$ and $g(r)$  in the metric provide quantum corrections at short scales,  ensuring regularity to our system.  The parameters $m$ and $Q$ represent the \textit{bare mass} \footnote{The bare mass $m$ is the mass-energy as seen by an observer who stands close to the origin. This mass contains no contribution from the electric energy and should not be confused  with the total electro-gravity mass $M$ (Arnowitt-Deser-Misner (ADM) mass) as seen by an observer at infinity.} and the electric charge of the source, respectively. The charge obeys the quantization rule $Q=N e\,$, where $N$ is a positive integer and $e$ denotes the elementary charge connected to the fine-structure constant $\alpha_{ e}$ through the relation $e^2= \alpha_{ e} \approx 1/137\,$.  For simplicity, we consider a positive charge for the black hole.

Proceeding to the solution of the Maxwell-like  equation \eqref{field_eqs} by using \eqref{F} and \eqref{line_el}, we get
\begin{equation} \label{eom1}
 E(r) = \f{Q}{r^2 \lagr_F}  \qquad \mathrm{and} \qquad \lagr_F = \f{\lagr'}{F'} = -\f{\lagr '}{E(r) E'(r)} \,.
\end{equation}
The prime symbol ($'$) denotes from now on differentiation with respect to the radial coordinate $r$. The component form of the Einstein field equations \eqref{field_eqs} is given by
\begin{align} \label{eom2}
\f{2m}{r^2} h'(r) + \f{Q^2 }{r^4} \left( g(r) - r g'(r)  \right) & = 8 \pi   \left( \rho_{\rm m}(r) + \rho_{\rm e}(r) \right) \,, \\ \label{eom2b}
- \f{  m}{r} h''(r) + \f{Q^2 }{r^4} \left( g(r) - r g'(r) + \f{r^2 g''(r)}{2} \right) & = 8\pi  \ p_{\bot}^{(\rm m)}(r) - 2 \lagr \,.
\end{align}
To solve the above differential equations, we need to specify the form of the matter density $\rho_{\rm m}(r)$ and the NED Lagrangian $\lagr\,$. Regarding the matter density, one may choose any regular  profile mentioned in the  literature. In our case, we will choose a matter density of the form
\begin{equation} \label{matter_d}
\rho_{\rm m}(r) = \f{3 \lpl^2 m}{8\pi  \left( r^2 + \lpl ^2/2 \right)^{5/2} } \,.
\end{equation}
According to \eqref{pressures}, the  density mentioned above produces radial and  tangential pressures, given  by
\begin{equation}
p_r^{(\rm m)}(r) = - \f{3 \lpl^2 m}{8\pi \left( r^2 + \lpl^2/2 \right)^{5/2} } \qquad \mathrm{and} \qquad p_{\bot}^{(\rm m)}(r) = 
\f{3\lpl^2 m \left( 3r^2 -\lpl^2 \right) }{16\pi \left(  r^2 + \lpl^2/2  \right)^{7/2} }  \,.
\end{equation}
The uncharged solution of the Einstein field equations  with the density \eqref{matter_d} results to a slightly modified Bardeen black hole \footnote{The Bardeen solution  has been derived uisng two different approaches; first, from the presence of a magnetic monopole subject to NED effects at short scales \cite{AyG00}, and second, from the corrected Newtonian potential, underlining the string T-duality and the path integral duality \cite{NSW19}. 
This is also a physical reason  to choose the density profile \eqref{matter_d}.} with metric potential of the form
\begin{equation} \label{Bar_metr}
f(r) = 1 - \frac{2  m \lpl^2  r^2}{ \left( r^2 + \lpl^2/2 \right)^{3/2} } \,.
\end{equation}
Contrary to the typical Bardeen solution, the  black hole described by \eqref{Bar_metr} satisfies the condition of \textit{self-implementation}\footnote{The term  \textit{self-implementation} was first introduced in \cite{NiS14}. It implies that the minimal cut-off length is of the same size as the minimal radius of the extremal black hole, meaning $r_0=\vartheta=\lpl$ in our case.}. 

\begin{figure}[!h] %
\centering
\includegraphics[height=1.95in]{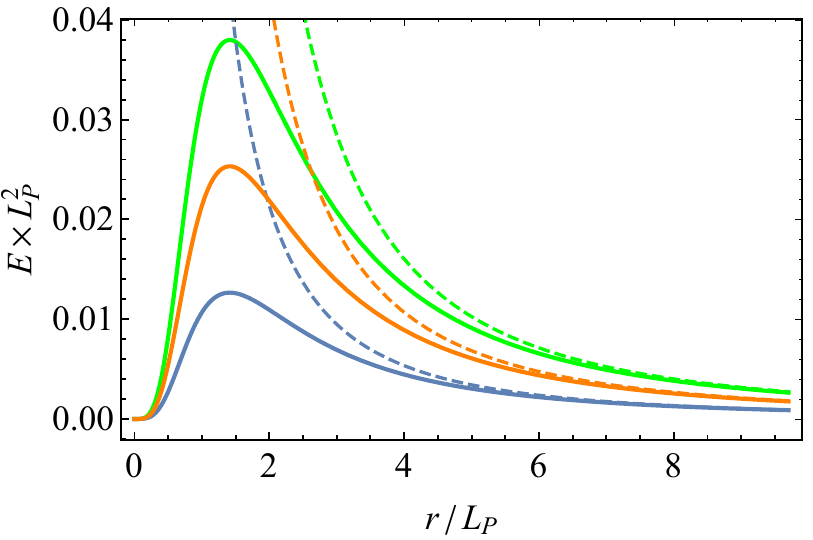}%
\caption{The NED electric field in Planck units. The blue line stands for $Q=e\,$, the orange line for $Q=2e$ and the green line for $Q=3e\,$. The dashed lines represent the classical divergent Coulomb fields for each case.} 
\label{fig:fig1}
\end{figure}

\begin{figure}[!h] %
\centering
\includegraphics[height=1.91in]{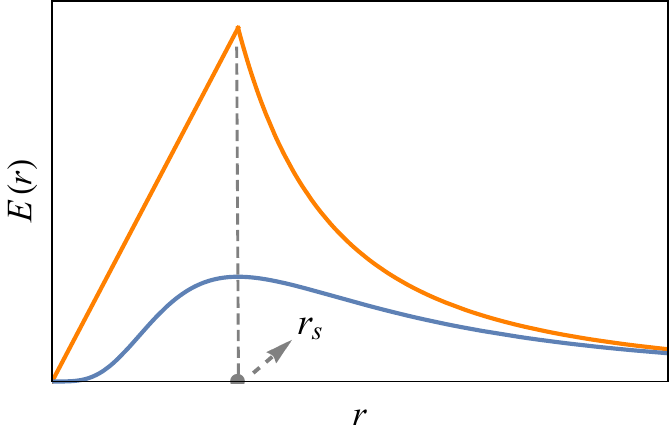}%
\caption{The classical  Coulomb field (orange line) and the field improved by NED (blue line) outside and inside of a  Planck-sized positively charged  sphere  with radius  $r_{\rm s}\,$.} 
\label{fig:fig2}
\end{figure}

\begin{figure}[!h] %
\centering
\subfigure[For fixed $Q=e$ and for $m=1.4 \mpl$ (blue line), for $m=m_0 \approx 0.915 \mpl$ (black dashed line) and  for $m=0.6 \mpl$ (gray line).] {%
\label{fig:fig3a}%
\includegraphics[height=1.72in]{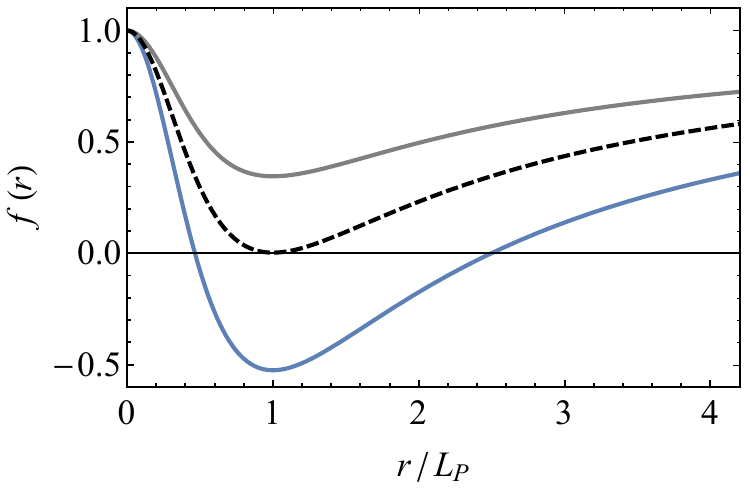}}%
\qquad
\subfigure[For fixed $m=\mpl$ and for $Q=e$ (black dashed line), for $Q=17e$ (orange line) and for $Q=20e$ (blue line).] {%
\label{fig:fig3b}%
\includegraphics[height=1.72in]{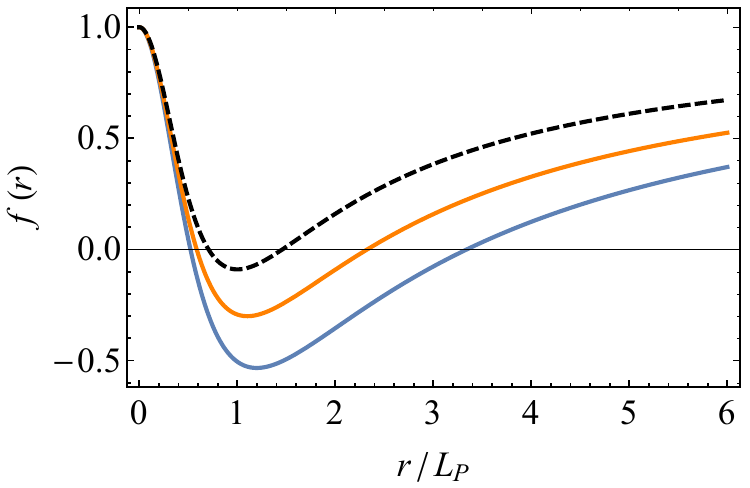}}%
\caption{The regular metric potential $f(r)$ \textit{vs} $r$ in Planck units.}
\label{fig:fig3}
\end{figure}

Regarding the electric sector, we need to find a suitable Lagrangian that satisfies the two initial limits and, simultaneously, yields a metric that is mathematically feasible. A suitable form is
\begin{equation} \label{lagrang}
\lagr = \lagr(F_{\rm M}) = - \f{Q^2 \left( \sqrt{-\f{Q^2}{2F_{\rm M}\lpl^4}} - 1 \right) }{2\lpl^4\left( \sqrt{-\f{Q^2}{2F_{\rm M}\lpl^4}} + 1 \right)^3} =  \f{Q^2}{2} \f{\lpl^2 - r^2 }{\left( r^2+ \lpl^2\right)^3 } \,,
\end{equation}
where the classical Maxwell field strength has the usual form of $F_{\rm M}=-Q^2/2r^4$. The  Lagrangian has a continues derivative 
\begin{equation} \label{d_lagrang}
\lagr ' = \f{2Q^2 r \left( r^2 - 2 \lpl^2 \right) }{\left( r^2 +  \lpl ^2 \right) ^4} \,.
\end{equation}
For weak fields ($r \gg \lpl $), the Lagrangian is approximately 
\begin{equation}
\lagr \approx -\f{Q^2}{2r^4} \left( 1 -  \f{4 \lpl^2}{r^2} + \mathcal{O}(r^{-4}) \right)  \approx -\f{Q^2}{2r^4} \equiv F_{\rm M}  \,,
\end{equation}
satisfying exactly the Maxwell theory. For strong fields ($r \ll  \lpl$), the Lagrangian has the regular form of
\begin{equation}
\lagr \approx \f{Q^2}{2 \lpl^4} \left( 1- \f{4 r^2}{ \lpl^2} +  \mathcal{O}(\lpl^{-4}) \right)  \approx \f{Q^2}{2 \lpl^4} = \rm{constant} .
\end{equation}
One could choose different Lagrangian functions that satisfy these limits; however, in our case, the chosen form is motivated by the string T-duality framework, which inspires charged black holes \cite{GaN22,GKN22}. In that context, a similar—but not identical—Lagrangian is derived, resulting in a regularized metric potential governed by an $\arctan$ function, as we will see below.
Having specified the expressions \eqref{matter_d} and \eqref{lagrang}, we can proceed to the solution of the field equations. Substituting  \eqref{lagrang} and \eqref{d_lagrang}   into \eqref{eom1},  we find the form of the electric field
\begin{equation} \label{el_field}
E(r) = \f{Q r^4}{\left( r^2 + \lpl ^2  \right)^3 } \,,
\end{equation}
as well as, the expressions
\begin{equation} \label{F_LF}
F= - \f{Q^2 r^8}{2\left(  r^2 +  \lpl^2 \right)^6 } \qquad \mathrm{and} \qquad \lagr_F = \f{\left( r^2 + \lpl^2 \right)^3 }{r^6} \,.
\end{equation}
In the classical regime ($r \gg  \lpl$), we recover the conventional Maxwell expressions:
\begin{equation}
E(r) \approx  \f{Q}{r^2} \,, \qquad F \approx - \f{Q^2}{2r^4} = F_{\rm M} \,, \qquad \lagr_F \approx 1 \,.
\end{equation}
The form of \eqref{el_field} is illustrated in Fig.~\ref{fig:fig1}.
NED cures the central divergence of the  Coulomb field by reaching the maximum  value of $E_{\rm max} = \f{4Q}{27\lpl^2}$ at $r_{\rm max}=\sqrt{2}  \lpl$  before decreasing monotonically and vanishing at $r=0\,$. The electric field \eqref{el_field} appears similarities with that of a homogeneously charged sphere with  radius  $r_{\rm s}=\sqrt{2}  \lpl\,$. In the case of the sphere, the classical field reaches its  maximum value on its surface  and then  decreases linearly from the inside, reaching zero at the center. However, in the classical sphere, there is a discontinuity on the surface. Here, NED has smoothed out such a discontinuity into a continuous behaviour. This allows to make a simple analogy: our source seems like a self-sustained, homogeneously charged,  Planck-sized sphere governed by NED. This analogy is plotted in Fig.~\ref{fig:fig2}.
In addition, the electric part of the stress-energy tensor \eqref{stress_tensors2} implies a charge density and two electric pressures that read
\begin{equation} \label{p_e}
\rho_e(r) = - p_r^{(\rm e)} (r) = \f{Q^2}{8\pi \left( r^2 +  \lpl^2 \right)^2 } \qquad \mathrm{and} \qquad p_{\bot}^{(\rm e)}(r) = \f{Q^2 \left( r^2 - \lpl^2 \right) }{8\pi \left( r^2 +  \lpl^2 \right)^3 } \,.
\end{equation}
Proceeding to the solution of the Einstein field equations \eqref{eom2} and \eqref{eom2b}, we obtain the following form of the regular metric potential 
\begin{equation} \label{rbh_metric}
f(r) = 1 - \frac{2  m \lpl^2  r^2}{ \left( r^2 + \lpl^2/2 \right)^{3/2} } + \f{Q^2 \lpl^2}{2 \left( r^2 +  \lpl^2 \right) } - \f{Q^2 \lpl}{2r} \arctan\left( \f{r}{ \lpl} \right)  .
\end{equation}
This implies that the two regular metric functions of \eqref{metric} are given by
\begin{equation}
h(r) = \f{r^3}{ \left( r^2 + \lpl^2/2 \right)^{3/2} } \qquad \mathrm{and} \qquad g(r)= \f{r^2}{2 \left( r^2+\lpl^2 \right) } - \f{r}{2 \lpl} \arctan \left( \f{r}{ \lpl} \right) .
\end{equation}
Now let us examine the behaviour of the black hole in different regimes:
\begin{itemize}
\item Near the origin ($r \ll \lpl$), the two functions are approximately given by
\begin{equation}
h(r) \approx  \f{2\sqrt{2} \ r^3}{ \lpl^3} + \mathcal{O}\left( \lpl^{-5} \right)   \qquad \mathrm{and} \qquad g(r) \approx - \f{r^4}{3\lpl^4} + \mathcal{O}\left( \lpl^{-6}  \right)
\end{equation}
and so the metric becomes
\begin{equation}
f(r) \approx 1 - \f{\Lambda_{\rm eff}}{3} r^2 \qquad \mathrm{with} \qquad \Lambda_{\rm eff} = \f{12 \sqrt{2} \ m \lpl + Q^2}{\lpl^2} > 0 \,.
\end{equation}
Therefore, the core resembles a repulsive de Sitter space and is regular everywhere, as both the Ricci and the Kretchmann scalar ($K=R_{\mu\nu\kappa\lambda} R^{\mu\nu\kappa\lambda}$) are  finite everywhere admitting respectively the central values of
\begin{equation}
R(0) =4 \Lambda_{\rm eff}  \qquad \mathrm{and} \qquad K(0) = \f{8}{3\lpl^4} \left( 228 m^2 \lpl^2 + 24\sqrt{2} \ m \lpl Q^2 + Q^4 \right) . 
\end{equation}
\item For $r \sim \lpl\,$, the black hole is described by the full NED potential \eqref{rbh_metric}.
\item For $r \gg \lpl\,$, the metric functions become
\begin{equation}
h(r) \approx 1 + \mathcal{O}(r^{-2})   \qquad \mathrm{and} \qquad g(r) \approx 1 - \f{\pi r}{4\lpl}  + \mathcal{O}(r^{-2})
\end{equation}
and the black hole coincides with the Reissner-N\"ordstrom metric
\begin{equation}
f(r) \approx 1 - \f{2M \lpl^2}{r} + \f{Q^2 \lpl^2}{r^2} \,.
\end{equation}
The mass $M$ is the total electro-gravity mass (ADM mass) consisting of 2 parts; the bare  mass $m$ and the electric energy-mass $m_{\rm e}$ resulting from the contribution of the charge  at infinity. The ADM mass can be derived from the Komar integral \cite{Kom59}, which takes the following form in our case
\begin{equation} \label{ADM_mass}
M = 4\pi \int\limits_{0}^{\infty} \d r \ r^2 \left( \rho_{\rm m} + \rho_{\rm e} \right) = m + m_{\rm e}  \qquad \mathrm{with} \qquad m_{\rm e}=\f{\pi Q^2}{8\lpl}\,.
\end{equation}
At this point, one may express the metric potential with respect to the total mass $M$, instead of the bare mass, as
\begin{equation} \label{metric_M}
f(r) = 1 -\f{2M \lpl^2}{r} h(r) + \frac{Q^2 \lpl^2}{r^2} \mathcal{G}(r) \,,
\end{equation}
where
\begin{equation}
\mathcal{G}(r) = \f{\pi r}{4 \lpl} h(r) + g(r) \,.
\end{equation}
\item  For even bigger distances ($r \ggg \lpl $),  only the contribution from the Schwarzschild part survives
\begin{equation}
f(r) \approx 1 - \f{2M \lpl^2}{r} \,.
\end{equation}
\item Asymptotically ($r \rightarrow \infty$) our metric obeys the Minkowski limit ($f(r) \rightarrow 1$).
\end{itemize}
Next, we study the horizon structure of the black hole for a fixed charge or a fixed bare mass, as depicted in Fig.~\ref{fig:fig3}.
 The increase in the mass or the charge, results in an increase in the event horizon. For a fixed charge, we get 3 different cases:
\begin{itemize}
\item For $m > m_0\,$, there are two horizons, namely a Cauchy $r_-$ (inner) and an event horizon $r_+$ (outer).
\item For $m=m_0\,$,  a degenerate horizon  is present with the smallest black hole mass $m_0$ and radius   $r_0=r_-=r_+\,$.
\item For $m<m_0\,$, no horizons exist, as the total energy of the system is  insufficient to form a black hole.
\end{itemize}
In Fig.~\ref{fig:fig3} we observe that for the smallest elementary charge $e$,  the extremal  black hole corresponds to $m_0 \approx 0.915 \mpl\,$, setting a lower bound for the mass required for the existence of horizons. Therefore, for  $m \geq 0.915 \mpl\,$, event horizons always exist for every value of the charge.

\begin{figure}[!h] %
\centering
\subfigure[ The Hawking temperature.] {%
\label{fig:fig4a}%
\includegraphics[height=1.72in]{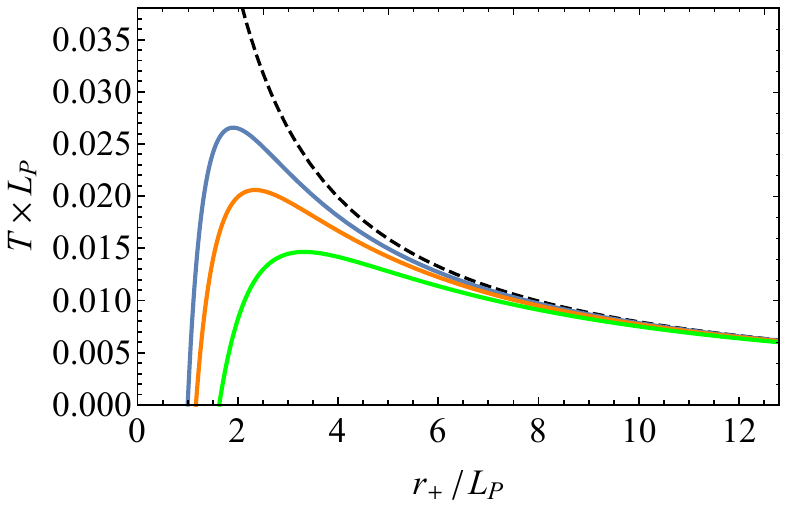}}%
\qquad
\subfigure[ The heat capacity.] {%
\label{fig:fig4b}%
\includegraphics[height=1.72in]{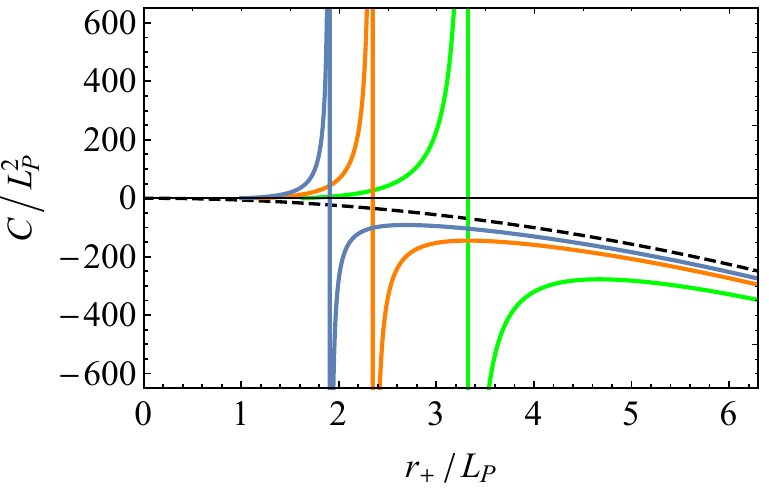}}%
\caption{The plots are illustrated  in Planck units, for  $Q=e$ (blue lines), for $Q=15e$ (orange lines) and for $Q=25e$ (green lines). The uncharged non-singular case ($Q=0$) almost coincides  with the blue lines, while the black dashed line corresponds to the semi-classical temperature of the Schwarzschild black hole.}
\label{fig:fig4}
\end{figure}

We shift our interest to the thermodynamic behaviour of the regular black hole by calculating the Hawking temperature \cite{Haw74} of the event horizon, which is proportional to its surface gravity and obeys the relation $T=  \f{1}{4\pi} \f{\d f(r)}{\d r} \bigg|_{r=r_+}$. The temperature is given by
\begin{align}  \nonumber
T =& \f{1}{8\pi r_+^2 (r_+^2 + \lpl^2)^2 (2r_+^2+\lpl^2)} \Big[3\lpl^3 Q^2 (r_+^2 + \lpl^2)^2 \arctan(r_+/\lpl)  \\ \label{Hawk_T}
&- r \left( \lpl^6 (4+3Q^2) + r_+^2 \lpl^4 (4+5Q^2)+4r_+^4\lpl^2 (Q^2-1)-4r_+^6 \right) \Big] \,
\end{align}
and its form is plotted in Fig.~\ref{fig:fig4a}. 
The temperature reaches a maximum before cooling down to a zero-temperature extremal  configuration, in contrast to the ultraviolet divergent temperature of the Schwarzschild black hole. The implementation of the minimal length has cured this pathology. This improvement is also a characteristic of all the other regular models mentioned in the introduction. The maximum of the temperature indicates a phase transition  between a small stable remnant and a large unstable black hole. The  transition  corresponds to a divergence of the heat capacity, which is given by the relation $C=\f{\d M}{\d T}\,$.  For the calculation of $C\,$,  the expression of the ADM mass $M$ is used in the derivative, since it coincides with the total internal energy of the system. Plugging \eqref{ADM_mass} in \eqref{rbh_metric} and using the horizon equation $f(r_+)=0\,$, we get
\begin{equation} \label{ADM_mass_v2}
M = \f{\pi Q^2}{8\lpl} + \f{r_+^2 + Q^2 \lpl^2 g(r_+)}{2r_+ \lpl^2  h(r_+)} = m_{\rm e}+m \,.
\end{equation}
The full expression of $C$ is omitted here, but it is depicted in Fig.~\ref{fig:fig4b}.
The thermally stable branch is governed by a positive $C\,$, whereas the unstable branch is associated with a negative $C$ where the  black hole radiates  continuously  until it reaches a cold,   stable remnant. Moreover, working in a canonical ensemble ($Q=\rm{fixed}$), we can derive the black hole entropy $S$ from the first law of black hole mechanics \cite{BCH73}, i.e., $\d M=T \d S\,$, retrieving
\begin{equation} \label{entropy}
S = \int\limits_{\lpl}^{r_+} \d r \ \f{2\pi r}{h(r)} = \pi r_+^2 \left[ \left( 1 - \f{\lpl^2}{r_+^2} \right)\sqrt{1+\f{\lpl^2}{r_+^2}}+\f{3\lpl^2}{2r_+^2} \ln\left( \f{2r_+ + \sqrt{4r_+^2+2\lpl^2}}{(2+\sqrt{6})\lpl} \right)   \right] .
\end{equation}
At large distances, the entropy \eqref{entropy} coincides with the area law up to some logarithmic corrections, while for $r_+=\lpl$ the entropy vanishes, being this way in agreement with the third law of thermodynamics.

\begin{figure}[!h] %
\centering
\includegraphics[height=2in]{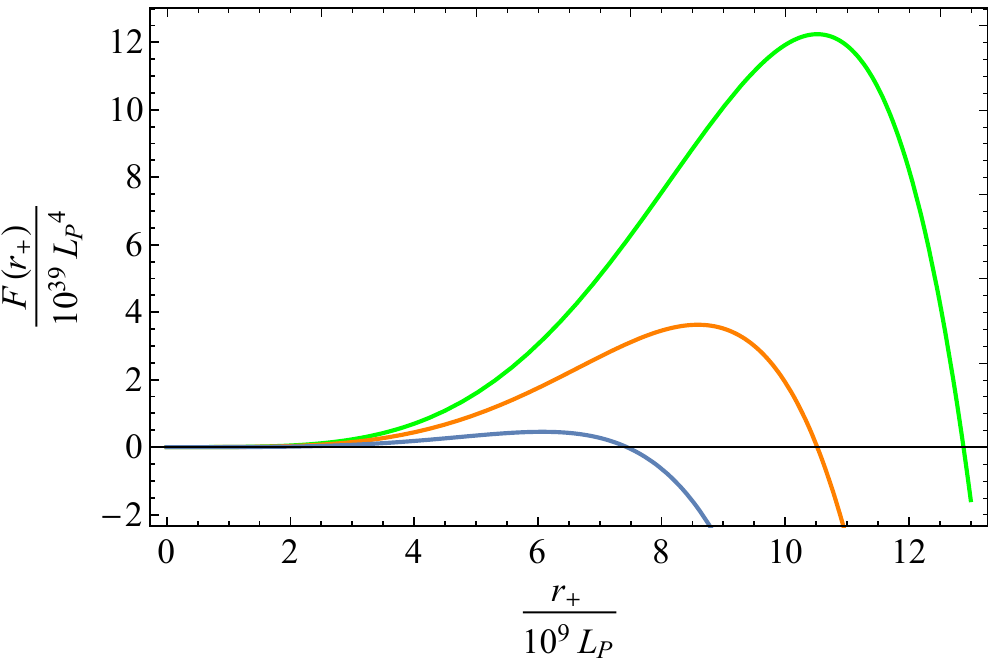}%
\caption{The function $F(r_+)$ \textit{vs} $r_+$ in Planck units for $N=1$ (blue line), for $N=2$ (orange line) and for $N=3$ (green line).} 
\label{fig:fig5a}
\end{figure}

We proceed by examining the four energy conditions \cite{MaV17} for both the matter and electric components of the total stress-energy tensor \eqref{stress_tensors1}:
\begin{enumerate}
\item Null energy condition: 
\begin{equation}
 \rho +   p_r \geq 0 \qquad \mathrm{and} \qquad  \rho   +   p_{\bot} \geq 0 \,,
\end{equation}
which are always true for each sector separately.
\item Weak energy condition: 
\begin{equation}
 \rho(r)  \geq 0\,,
\end{equation}
which is true for both densities.
\item Dominant energy condition: 
\begin{equation}
 \rho(r) \geq  | p_{\rm r}(r) |  \qquad \mathrm{and} \qquad  \rho(r) \geq |  p_{\bot}(r) |\,.
\end{equation}
This condition is satisfied at all distances if the combined densities from both sectors are considered, i.e., $\rho_{\rm m}(r)+ \rho_{\rm e}(r) \geq |  p_{\bot}^{(\rm m)}(r) + p_{\bot}^{(\rm e)}(r)|$. In this case, the mass must be fixed, and the condition is met for charge values above a certain threshold relative to the mass. 
\item  Strong energy condition: 
\begin{equation}
 \rho(r) + p_{\rm r}(r) + 2 p_{\bot}(r)  \geq 0\,.
\end{equation}
This condition is meaningful for the matter part of the stress-energy tensor and is valid for $r \geq \lpl / \sqrt{3} \,$. Within this range, gravity maintains its attractive character, while for smaller distances, gravity becomes repulsive (de Sitter core). Unfortunately,  it is necessary to violate the strong energy condition near the origin to prevent singularities from forming and to maintain ultraviolet self-completeness. Nonetheless, the energy condition is preserved outside the Planck sphere, which represents the physically meaningful region.
\end{enumerate}
Last but not least, let us offer a qualitative insight into the discharge process of this quantum-corrected black hole through the Schwinger effect \cite{Sch51}. According to this mechanism, electron-positron pairs can be produced just outside the event horizon  in the presence of a strong electric field. One particle of the pair will be emitted while the other will be absorbed, leading to the decay of the black hole's electric charge. For this phenomenon to occur, the electric field \eqref{el_field} of the black hole must exceed a critical value $E_{\rm cr}=\pi m_{e}^2/e\,$, resulting in the following relation
\begin{equation}
\frac{N e r_+^4}{(r_+^2+\lpl^2)^3} \geq \pi \frac{m_e^2}{e} \,,
\end{equation}
where we have used the quantization of the charge $Q=Ne\,$. The critical value $E_{\rm cr}$ results from quantum field theory calculations in Minkowski space. Although corrections are expected in curved spacetime \cite{Gas83}, we adopt $E_{\rm cr}=\pi m_{e}^2/e$ in our calculations, as it is anticipated to provide the dominant contribution to our qualitative analysis.   The electron mass $m_e$ is approximately $m_e \approx  10^{-23} \mpl$, and thus the above inequality becomes
\begin{equation} \label{Fr}
F(r_+) \geq 0 \qquad \mathrm{with} \qquad F(r_+) =  N r_+^4- \frac{\pi m_e^2}{e^2} (r_+^2+\lpl^2)^3 \,.
\end{equation}
The positivity of the function $F(r_+)$ indicates the onset of Schwinger pair production at the event horizon. For this reason, we illustrate its form for various charge numbers in Fig.~\ref{fig:fig5a}.
Given the extremely small value of the electron mass relative to the Planck mass, the second term in \eqref{Fr} becomes negligible in comparison with the first term, resulting in $F(r_+)\geq 0$ unless the event horizon becomes exceedingly large compared to the Planck scale, thereby compensating for the electron mass term.  For instance, if $N=1$ then $F(r_+)\geq 0$ for $r_+ \leq 7.44 \cdot 10^9 \lpl$, while for $r_+ > 7.44 \cdot 10^9 \lpl$ the function becomes negative ($F(r_+)<0$). Consequently, for galactic size black holes, the electric field is not strong enough to produce a Schwinger pair $\pm e\,$, whereas for Planck-sized objects, the field is always sufficient to nucleate the corresponding pair regardless of the charge value.

\section{Cosmological anti-de Sitter solution}
\label{sec:ads_rbh}

\begin{figure}[!h] %
\centering
\includegraphics[height=1.95in]{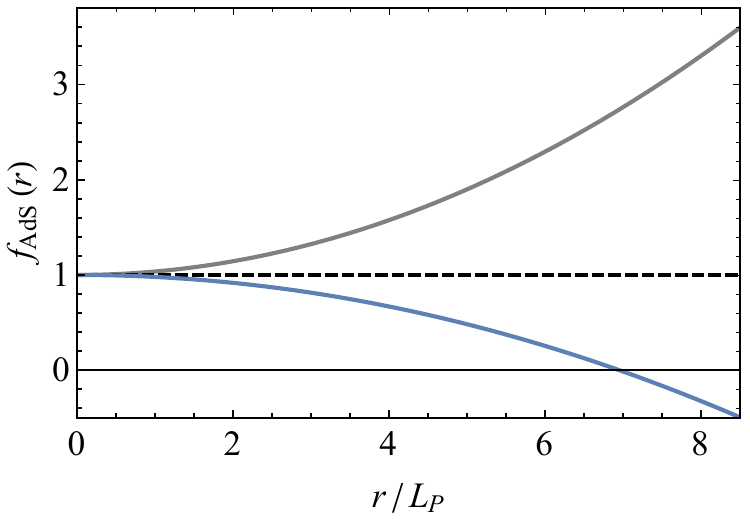}%
\caption{The quadratic form of the metric potential near the origin in Planck units for fixed  $N_{\rm min}=1$ and $\ell=\lpl $ and for $m=0.18\mpl$ (blue curve), for $m=\mathcal{M}_{\rm max} \simeq 0.176 \mpl$ (black dashed line) and for $m =0.17\mpl $ (gray line).} 
\label{fig:fig5}
\end{figure}

In the preceding section, we derived an asymptotically flat, non-singular black hole solution. This solution is ultraviolet improved by incorporating the Planck length as a renormalization scale with additional corrections arising from NED at distances near this scale. In this section and the following one, we extend the aforementioned solution within a cosmological background governed by $\Lambda$. We start the discussion with AdS, introducing a negative $\Lambda$ into the action \eqref{action}. The Einstein field equations  provide a metric potential of the following form
\begin{equation} \label{ads_metric1}
f_{\rm AdS}(r) = f(r) + \frac{r^2}{\ell^2} \,,
\end{equation}
where $f(r)$ is given by \eqref{rbh_metric} and $\ell$ is the AdS radius connected with $\Lambda$ through the relation $\Lambda = - 3/\ell^2 $. The term $r^2/\ell^2$  in the potential \eqref{ads_metric1} is positive,  implying that gravitational attraction increases with distance. This leads to a contracting black hole universe. At large distances, the above line element coincides with the classical Reissner-N\"ordstrom-AdS solution, recovering asymptotically an empty AdS space. Near the origin, we obtain
\begin{equation} \label{ads_metric2}
f_{\rm AdS} (r) \approx 1 - \f{\Lambda_{\rm eff}}{3} r^2 \qquad \mathrm{with} \qquad \Lambda_{\rm eff} =  \f{12 \sqrt{2} \ m \lpl + Q^2}{\lpl^2} - \f{3}{\ell^2} 
\end{equation}
and so we have three different cases:
\begin{enumerate}
\item For $m > \f{3\lpl^2-\ell^2 Q^2}{12\sqrt{2} \ell^2 \lpl} = \mathcal{M}\,$, the core is repulsive with $\Lambda_{\rm eff} >0\,$ (de Sitter core).
\item For $m = \mathcal{M}\,$, the core corresponds to Minkowski space and is free of gravity since $\Lambda_{\rm eff}=0\,$ (Minkowski core).
\item For $m < \mathcal{M}\,$, the core is attractive with $\Lambda_{\rm eff}<0\,$ (AdS core).
\end{enumerate}
The 2nd and the 3rd cases, involving a Minkowski and an  AdS core, signal instability \cite{NiT11}. However, these cases will not bother us because we will show  that for such values of the bare mass, no black hole horizons exist.
Firstly, if the numerator of $\mathcal{M}$  is negative ($3\lpl^2-\ell^2 Q^2 < 0$), then the 1st condition is always satisfied, leading to a repulsive core. This is because the bare mass $m$ is considered to be positive a \textit{priori}. Secondly, if the numerator of $\mathcal{M}$ is non-negative ($3\lpl^2-\ell^2 Q^2 \geq 0$), the AdS radius satisfies the condition $\ell^2 \leq 3\lpl^2/Q^2$. Considering the charge quantization, this condition becomes $\ell^2 \leq 411 \lpl^2/N^2 $. This inequality restricts the integer $N$ to be $N \leq 20\,$, otherwise the radius $\ell$ would penetrate  the forbidden sub-Planckian regime. 
Let us consider, for example, the lowest theoretical value of the AdS radius, $\ell_{\rm min}=\lpl \,$, which actually implies the inequality $1 \leq N \leq 20$ for having $\mathcal{M} \geq 0\,$. In this case, the maximum  allowed charge  is for $N_{\rm max} = 20$ giving   $\mathcal{M}_{\min} \simeq 0.0047 \mpl\,$, while the minimum is for $N_{\min}=1$ giving $\mathcal{M}_{\max} \simeq 0.176 \mpl\,$. 
The  quadratic form of the metric \eqref{ads_metric2}  is depicted in Fig.~\ref{fig:fig5} for the above pair ($N_{\min},\mathcal{M}_{\rm max}$). 
As observed from Fig.~\ref{fig:fig5}, if $m \leq \mathcal{M}_{\rm max} \,$, which is actually the condition for having an AdS or Minkowski core, the curves of the potential never intersect the $r$-axis, indicating this way the non-existence of horizons. The same phenomenology holds if we consider the pair ($N_{\max},\mathcal{M}_{\rm min}$) instead.

\begin{figure}[!h] %
\centering
\includegraphics[height=1.95in]{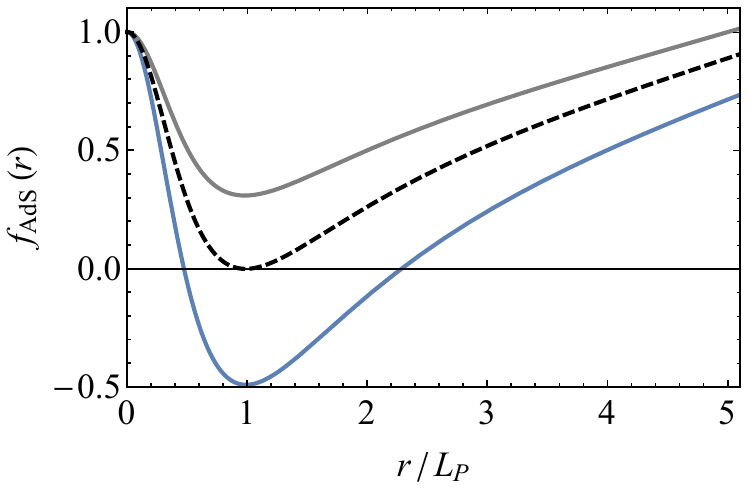}%
\caption{The AdS metric potential in Planck units for fixed $\ell=10\lpl$ and $Q=e$ and for $m=1.7\mpl$ (blue line),  for $m=m_0 \approx 1.01\mpl$ (black dashed line) and for $m=0.7\mpl$ (gray line).} 
\label{fig:fig6}
\end{figure}

We move on to examine  the horizon structure of the spacetime \eqref{ads_metric1} by plotting all possible cases in Fig~\ref{fig:fig6}.
The horizon topology is the same with the asymptotically flat case featuring two, one or no horizons depending on the value of the bare mass $m$ relative to an extremal value $m_0$. AdS space represents a hyperbolic manifold  with  negative  curvature, resembling the geometry of a torus. Hence, the cosmological horizon is absent from the observable universe, as it extends to spatial infinity for a timelike observer. 

\begin{figure}[!h] %
\centering
\includegraphics[height=1.95in]{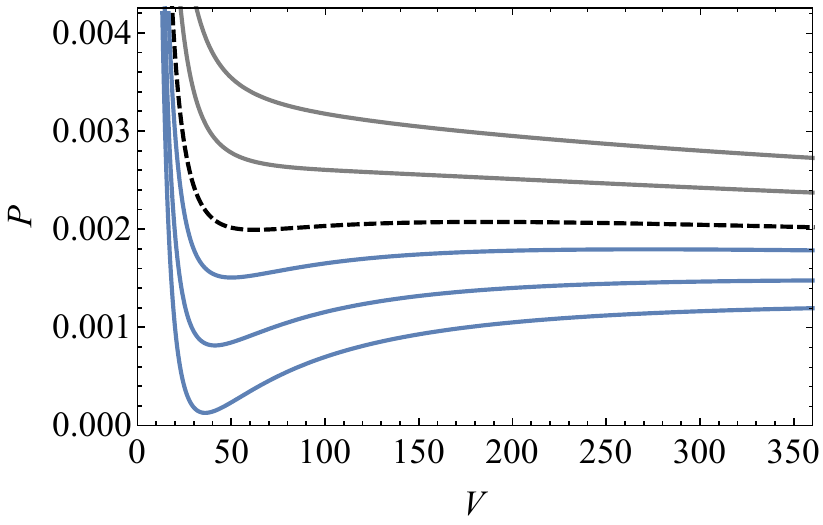}%
\caption{The isotherms in the $P-V$ plane for $Q=e$ and $\lpl=1\,$. The black dashed line stands for the critical temperature $T_{\rm c}\simeq 0.034\,$, the blue lines stand for $T<T_{\rm c}$ and the gray lines for $T>T_{\rm c}\,$.} 
\label{fig:fig7}
\end{figure}

\begin{figure}[!h] %
\centering
\subfigure[The AdS black hole temperature.] {%
\label{fig:fig8a}%
\includegraphics[height=1.72in]{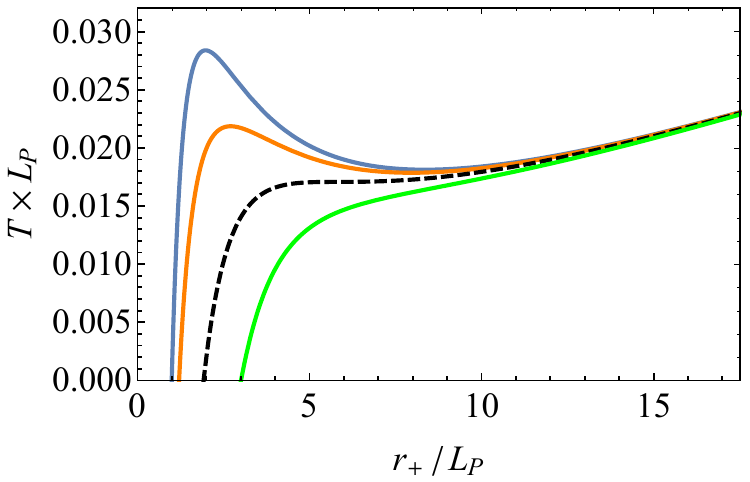}}%
\qquad
\subfigure[The heat capacity $C_P\,$.] {%
\label{fig:fig8b}%
\includegraphics[height=1.72in]{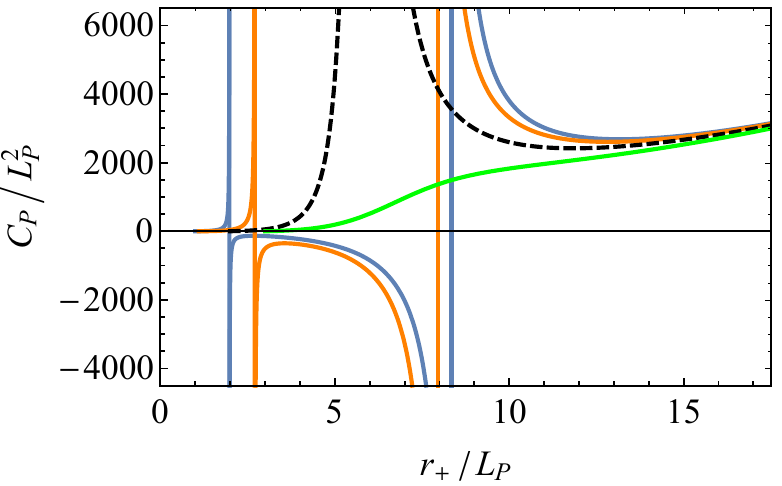}}%
\caption{The plots are illustrated in Planck units, for fixed $\ell=15\lpl$ and for $Q=e$ (blue lines), for $Q=17e$ (orange lines), for $Q=Q_0=30e$ (black dashed lines) and for $Q=45e$ (green lines). }
\label{fig:fig8}
\end{figure}

The next interesting aspect concerns the thermodynamic properties of this spacetime. At this point, we take a step further and extend our phase space in accordance with the proposal of black hole chemistry \cite{KuM17}. This extension allows us to  connect  black hole systems with real known chemical systems, such as Van der Waals fluids \cite{KuM12,RKM14,Tzi19}, reentrant phase transitions of nicotine/water \cite{AKM13,FKM14} and triple points like water \cite{AKM+14}. Such identifications become possible because  the cosmological constant is treated as a thermodynamic variable, playing the role of the  pressure $P$, which is given by the relation
\begin{equation} \label{ads_press}
P = \f{3}{8\pi \lpl^2 \ell^2} \,.
\end{equation}
The introduction of $P$  also implies the introduction of its conjugate quantity $V$ that plays the role of thermodynamic volume \cite{Dol12,Dol+11}. This setup enables us to build an equation of state $P=P(V,T)$ for the black hole and compare it with the equation of state of well-known chemical systems.
In addition, the total energy-mass of the AdS black hole no longer represents the internal energy. Instead, the mass is interpreted as  the chemical enthalpy $H$ \footnote{The ADM mass $M$ represents the internal energy $U$ that is needed to create a black hole plus the  pressure-volume energy  needed to place it inside the cosmological background \cite{KRT09,CGK++11}.}  given by  
\begin{equation}
M \equiv H = U +PV \,.
\end{equation}
The ADM mass $M$ can be defined at infinity and is given again by \eqref{ADM_mass}. Substituting the bare mass $m$ from \eqref{ADM_mass} into \eqref{ads_metric1}, we obtain the AdS metric potential  in terms of the total energy 
\begin{equation} \label{ads_metric3}
f_{\rm AdS}(r) = 1 - \f{2 }{r} \left( M - \f{\pi Q^2}{8 \lpl} \right)h(r)  + \f{Q^2\lpl^2}{r^2} g(r) + \f{r^2}{\ell^2} \,.
\end{equation}
From the horizon equation $f_{\rm AdS}(r_+)=0\,$, we get
\begin{equation}
M =  \f{\pi Q^2}{8\lpl} + \f{\ell^2 r^2_+ + r_+^4 + \ell^2 Q^2\lpl^2 g(r_+)}{2\ell^2 \lpl^2 r_+ h(r_+)} = m_e + m \,,
\end{equation}
which allows us to calculate the thermodynamic volume as
\begin{equation} \label{ads_vol}
V = \left( \f{\p M}{\p P} \right)_{S,Q} =  \f{4\pi r_+^3}{3h(r_+)} = \f{4\pi}{3} (r_+^2+\lpl^2/2)^{3/2} .
\end{equation}
In the classical regime, the thermodynamic volume coincides with that of a Euclidean sphere, even though it does not posses a geometric interpretation as an  actual sphere. One must not confuse the thermodynamic volume of the black hole with its actual geometric volume. Furthermore, the temperature resulting from the surface gravity is give by
\begin{equation} \label{ads_temp}
T= \f{1}{4\pi r_+} \left( 1 + \f{3r^2_+}{\ell^2} - \f{Q^2\lpl^2}{r_+^2} g(r_+) + \f{Q^2 \lpl^2}{r_+} g'(r_+) - \left( 1 + \f{r_+^2}{\ell^2} + \f{Q^2\lpl^2}{r_+^2} g(r_+) \right) \f{r_+ h'(r_+)}{h(r_+)}  \right) .
\end{equation}
Solving \eqref{ads_vol} and \eqref{ads_press} with respect to
\begin{equation}
r_+ = \f{1}{2} \sqrt{\left( \f{6V}{\pi}\right)^{2/3}-2\lpl^2 } \qquad \mathrm{and} \qquad \ell^2 = \f{3}{8\pi \lpl^2 P}
\end{equation}
and plugging these expressions in the temperature \eqref{ads_temp}, we retrieve an equation of state $P=P(V,T)$ for the AdS black hole. The explicit form of this equation is rather lengthy and is therefore omitted. Nevertheless, we plot the isotherms in Fig.~\ref{fig:fig7}.
Above some critical temperature ($T>T_{\rm c}$), the isotherms are well-behaved as the increase of the volume is followed by a decrease in pressure, as expected. Below the critical temperature ($T<T_{\rm c}$), the isotherms exhibit an unphysical branch for the pressure, similar to that of a Van der Waals gas. This indicates the existence of a first order phase transition. This can be verified by examining the behaviour of the temperature \eqref{ads_temp}, as depicted in Fig.~\ref{fig:fig8a}.
Contrary to the temperature \eqref{Hawk_T} of the asymptotically flat case, the  AdS black hole temperature can have two, one or no extrema. For $Q>Q_0\,$, there exist  two extrema corresponding to a black hole phase transition, while for  $Q<Q_0\,$, the system is thermally stable with no phase transition. When $Q=Q_0\,$, the two extrema merge at one inflexion point. 

The nature of the transition can be determined from the heat capacity. The appearance of a PV$-$term implies the existence of two heat capacities; one with constant volume $C_V=\left( \f{\p M}{\p T} \right)_{V} =0\,$, which is always zero in our case, and one with constant pressure $C_P=\left( \f{\p M}{\p T} \right)_{P}\,$, which is not zero and is illustrated in Fig.~\ref{fig:fig8b}.
For $Q<Q_0\,$, there are two divergences in $C_P\,$, dividing the space into a small stable, a middle unstable and a large stable branch. This can be identified as a first order phase transition between a small/large stable black hole, reminiscent of the liquid/gas transition of a Van der Waals gas. This is in contrast to the conventional Hawking-Page transition \cite{HaP83} of its classical AdS counterpart. For $Q=Q_0$ there is only one divergence in $C_P$ where the two stable black holes coexist at the inflexion point, while for $Q>Q_0$ the heat capacity is always positive with no indication for a phase transition.

\section{Cosmological de Sitter solution}
\label{sec:ds_rbh}

In this section, we explore the geometrical properties of the regular black hole within a cosmological background governed by a positive cosmological constant $\Lambda >0\,$. This spacetime, known as de Sitter, characterizes an expanding black hole universe and has a metric potential of the form
\begin{equation} \label{ds_metric}
f_{\rm dS}(r) = f(r) - \f{\Lambda}{3} r^2  .
\end{equation}
The potential of the $\Lambda-$term decreases with the increase of the distance, resulting in an anti-gravitational effect similar to that of dark energy. This behaviour mimics the accelerated expansion of the universe.  At large distances, the black hole approximates the classical Reissner-N\"ordstrom-de Sitter solution, and asymptotically,  the spacetime resembles empty de Sitter space. Near the origin,  the metric potential becomes
\begin{equation}
f_{\rm dS}(r) \approx  1 - \f{\Lambda_{\rm eff}}{3} r^2 \qquad \mathrm{with} \qquad \Lambda_{\rm eff}  =  \f{12 \sqrt{2} \ m \lpl + Q^2}{\lpl^2} + \Lambda >0 \,,
\end{equation}
implying always the existence of a repulsive de Sitter core, pushing against gravitational collapse. The horizon structure is richer than in previous cases, featuring up to three apparent singularities; a Cauchy $r_-\,$, an event $r_+$ and a cosmological horizon $r_{\rm c}\,$. In the presence of distinct horizons, an observer standing between the event and the cosmological horizon experiances two different Hawking temperatures because the surface gravity differs on the two horizons. Consequently, the region between them is not in thermal equilibrium. 
The only way to achieve equilibrium is by equalizing the temperatures of the radiating horizons.  
This can be achieved through a unique relationship between the black hole parameters  or by having an extremal black hole  where there is a horizon degeneracy \cite{MaR95}. The cases providing thermal equilibrium are commonly explored within the context of Euclidean quantum gravity  upon Wick-rotating the real time axis $t$ to an imaginary time $\tau$ through the relation $\tau=it\,$. Further elaboration on this concept  will be provided in the next section. The associated manifolds are commonly known as gravitational instantons\footnote{A gravitational instanton is a non-singular Riemanian manifold with a Euclidean signature (++++), satisfying the Euclidean version of the Einstein field equations \cite{GiH79}.}.  In the following subsections we examine four different gravitational instantons.  

\subsection{Lukewarm black hole}
\label{subsec:luke}

\begin{figure}[!h] %
\centering
\includegraphics[height=1.95in]{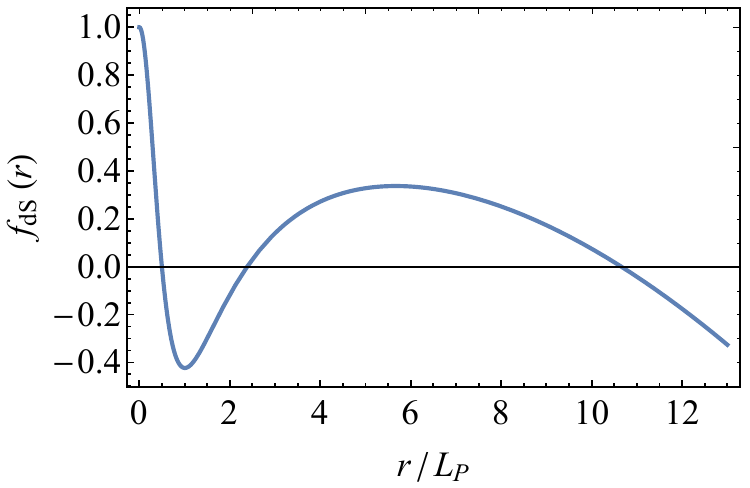}%
\caption{The lukewarm metric potential in Planck units for $Q=e\,$, $\Lambda=0.02  \lpl^{-2}$ and $m=1.3\mpl\,$.} 
\label{fig:fig9}
\end{figure}

The lukewarm black hole \cite{Rom92} is a spacetime with three distinct horizons, as illustrated in Fig.~\ref{fig:fig9}. 
We rename the two radii as $r_+=a$ and $r_{\rm c}=b\,$. Since the two horizons are different, thermal equilibrium can only be achieved through a proper relation between the bare mass,  the charge and the cosmological constant.  This relation can be determined by solving the following system
\begin{equation} \label{lw_eq}
f_{\rm dS}(a)=f_{\rm dS}(b)=0 \qquad \mathrm{and} \qquad f'_{\rm dS}(a)=-f'_{\rm dS}(b) \,.
\end{equation}
The left part of \eqref{lw_eq} is the horizon equation, while the right part implies that the two surface gravities are equal but with opposite sign, matching this way the two Hawking temperatures  $T_a=T_b\,$.  Solving the left equation   \eqref{lw_eq}, we get
\begin{equation} \label{lw_m}
m=m(a,b,\Lambda)= \f{b^2 (b^2 \Lambda - 3)g(a)+a^2 (3-a^2 \Lambda)g(b)}{6\lpl^2 \left[ ah(a) g(b) - b h(b) g(a)  \right] }
\end{equation}
and
\begin{equation} \label{lw_Q}
Q^2=Q^2(a,b,\Lambda) = \f{ab \left[  b (b^2 \Lambda-3)h(a) + a  (3 - a^2 \Lambda ) h(b) \right] }{3\lpl^2 \left[ ah(a) g(b) - b h(b) g(a)  \right] } \,.
\end{equation}
Plugging the expressions \eqref{lw_m} and \eqref{lw_Q} in the right equation \eqref{lw_eq}, we get
\begin{equation} \label{lw_rel}
m=m(a,b) \, \qquad Q^2=Q^2(a,b) \qquad \mathrm{and} \qquad \Lambda=\Lambda(a,b) \,.
\end{equation}
The relations \eqref{lw_rel} are omitted here due to their lengthy form, but they imply constrains on the  distance between the two horizons to ensure positivity for the lukewarm parameters ($m,Q^2,\Lambda$).
The lukewarm intanton is then expressed as
\begin{equation}
\d s^2 = f_{\rm lw}(r) \d \tau^2 + f_{\rm lw}(r)^{-1} \d r^2 + r^2\d \Omega^2 \,,
\end{equation}
where $\d \Omega^2= \d \theta^2 + \sin^2 \theta \d \phi^2$ is the line element of a 2-sphere.

\subsection{Nariai black hole}
\label{subsec:nariai}

\begin{figure}[!h] %
\centering
\includegraphics[height=1.95in]{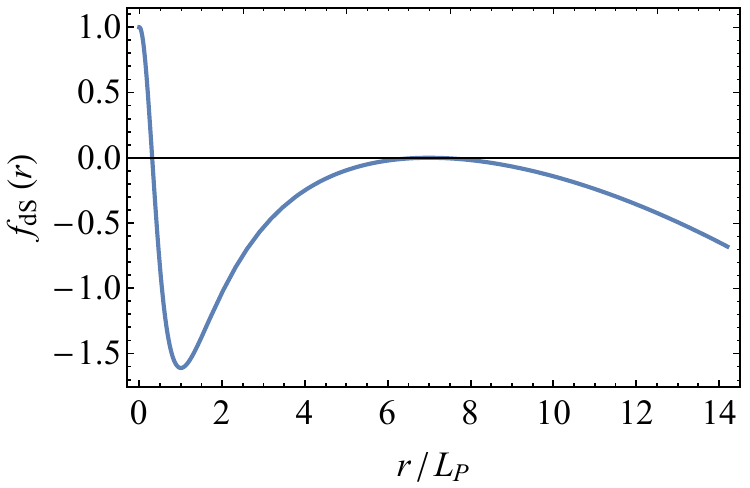}%
\caption{The Nariai metric potential in Planck units for $Q=e\,$, $\Lambda=0.02  \lpl^{-2}$ and $m\simeq 2.391\mpl\,$.} 
\label{fig:fig10}
\end{figure}

Nariai spacetime \cite{Nar99} represents the degenerate case where the event horizon is of the same size with the cosmological horizon ($r_+=r_{\rm c}=\varrho $), as illustrated in Fig.~\ref{fig:fig10}.
The degeneracy implies that the singularity lies at an infinite proper distance from any point to the degenerate horizon, leading to a vanishing Hawking temperature. Therefore, the Nariai black hole parameters can be found from the relation \cite{Bou97}
\begin{equation} \label{nar_eq}
f_{\rm dS}(\varrho)=f'_{\rm dS}(\varrho)=0 \,,
\end{equation}
which gives
\begin{equation} \label{nar_m}
m= \f{\varrho \left[ 2g(\varrho) (3-2\varrho^2 \Lambda)+\varrho (\varrho^2 \Lambda -3)g'(\varrho) \right] }{6\lpl^2 \left[ g(\varrho)(h(\varrho)+\varrho h'(\varrho)) - \varrho h(\varrho) g'(\varrho) \right] }
\end{equation}
and
\begin{equation} \label{nar_Q}
Q^2 = \f{\varrho^2 \left[ 3h(\varrho) (1-\varrho^2 \Lambda)+ \varrho h'(\varrho) (\varrho^2 \Lambda -3) \right] }{3\lpl^2 \left[ g(\varrho)(h(\varrho)+\varrho h'(\varrho)) - \varrho h(\varrho) g'(\varrho) \right]} \,.
\end{equation}
It has been shown that, for near extremal black holes, one may approximate the potential with respect to the horizons as \cite{Sha17}
\begin{equation}
f_{\rm{N}}(r) \approx \f{f''_{\rm dS}(\varrho)}{2} (r-r_+) (r-r_{\rm c}) \,,
\end{equation}
where $f_{\rm dS}''(\varrho)= \f{\d ^2 f_{\rm dS}(r)}{\d r^2} \Big|_{r=\varrho} $.  The function $f''_{\rm dS}(\varrho)$ is negative, as can be seen from the nature of the potential near the degenerate horizon (local maximum), maintaining this way a positive-definite metric. In the Euclidean section, the Nariai instanton reads
\begin{equation} \label{Nar_el}
\d s^2 = f_{\rm N}(r) \d \tau^2 + f_{\rm N}(r)^{-1} \d r^2 + r^2\d \Omega^2 \,,
\end{equation}
but these coordinates are inappropriate for this solution, as $f_{\rm N}(r) \rightarrow 0$ between the infinitesimal distance of the two horizons. We make the following coordinate transformation for near extremal black holes:
\begin{equation} \label{Nariai_trs}
\tau = -\f{2\xi}{\epsilon f''_{\rm dS}(\varrho)} \qquad \mathrm{and} \qquad r=\varrho - \epsilon \cos \chi \,,
\end{equation}
where $\epsilon$ is an arbitrary tiny length scale  and $\xi$ and $\chi$ are periodic variables with periods $2\pi$ and $\pi$ respectively.
For $\chi=0\,$, we get the black hole radius $r=\varrho-\epsilon=r_+\,$, while for $\chi=\pi\,$, we get the cosmological radius $r=\varrho +\epsilon=r_{\rm c}\,$. In the exact Nariai limit ($\epsilon \rightarrow 0$), the two horizons coincide. Then, the metric  potential becomes
\begin{equation} \label{Nar_pot}
f_{\rm N}(\chi) = - \f{\epsilon^2 f''_{\rm dS}(\varrho)}{2} \sin^2\chi \,.
\end{equation}
Substituting the new coordinates and the potential \eqref{Nar_pot} in \eqref{Nar_el} and taking the exact Nariai limit $\epsilon \rightarrow 0\,$, we retrieve the  Nariai line element
\begin{equation}
\d s^2 = -\f{2}{f''_{\rm dS}(\varrho)} \left( \d \chi^2 + \sin^2\chi \d \xi ^2 \right) + \varrho^2 \d \Omega^2 \,,
\end{equation}
which has a geometrical topology of a round 2-sphere  $ \rm \d S^2 \times S^2$.

\subsection{Cold black hole}
\label{subsec:cold}

\begin{figure}[!h] %
\centering
\includegraphics[height=1.95in]{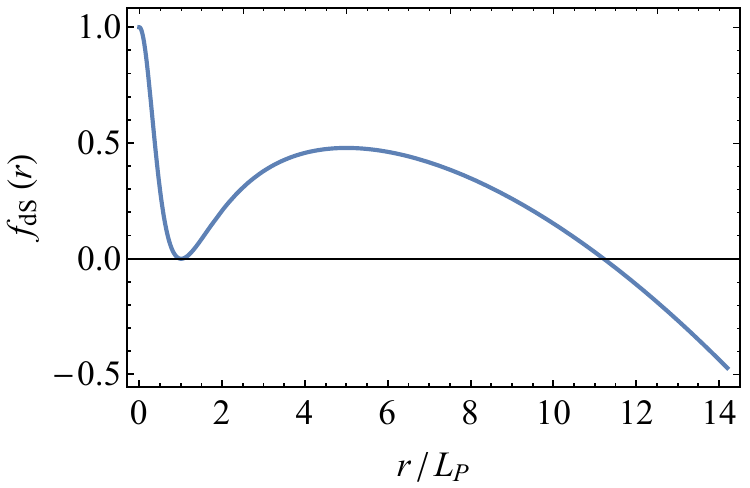}%
\caption{The cold metric potential in Planck units for $Q=e\,$, $\Lambda=0.02  \lpl^{-2}$ and $m\simeq 0.912\mpl\,$.} 
\label{fig:fig11}
\end{figure}

The cold black hole is another extremal solution where the Cauchy and the event horizon coincide $r_-=r_+=\varrho_{{\rm C}}\,$, as depicted in Fig.~\ref{fig:fig11}.
The degeneracy of the two horizons implies an infinite proper distance from any point to $\varrho_{{\rm C}}$ and, consequently, a vanishing Hawking temperature, leaving the   spacetime in between in thermal equilibrium with the radiation of the cosmological surface. We can remove the cosmological conical singularity through the periodicity of the Euclidean time $\tau$. As before, the extremal potential condition 
\begin{equation} \label{cold_rel}
f_{\rm dS}(\varrho_{\rm C})=f_{\rm dS}'(\varrho_{\rm C})=0 \,
\end{equation}
holds true, leading to similar relations for $m$ and $Q$ as in the Nariai case, but with different degenerate root $\varrho_{\rm C}\,$. In this case, $f_{\rm dS}''(\varrho_{\rm C})>0$ due to the local minimum and  the metric potential can be approximated by
\begin{equation}
f_{\rm C}(r) \approx \f{f''_{\rm dS}(\varrho_{\rm C})}{2} (r-r_-) (r_+-r) \,.
\end{equation}
Introducing the coordinates ($\xi,\chi$)  and performing the transformation \cite{MST13}
\begin{equation}
\tau = \f{2\xi}{\epsilon f''_{\rm dS}(\varrho)}  \qquad \mathrm{and} \qquad r=\varrho + \epsilon \cosh  \chi   \,,
\end{equation}
we can write the cold potential as
\begin{equation}
f_{\rm C}(\chi) = \f{f''_{\rm dS}(\varrho_{\rm C})}{2} \epsilon^2 \sinh^2\chi \,.
\end{equation}
Substituting these expressions in the Euclidean line element and taking the limit $\epsilon \rightarrow 0\,$, we obtain the  cold instanton
\begin{equation}
\d s^2 = \f{2}{f''_{\rm dS}(\varrho_{\rm C})} \left( \d \chi^2 + \sinh^2\chi \d \xi^2  \right) +  \varrho_{\rm C}^2  \d \Omega^2 \,,
\end{equation}
which resembles topologically  an extremal black hole in AdS spacetime ($\mathrm{AdS}^2 \times \rm S^2$).

\subsection{Ultracold black hole}
\label{subsec:ultracold}

\begin{figure}[!h] %
\centering
\includegraphics[height=1.95in]{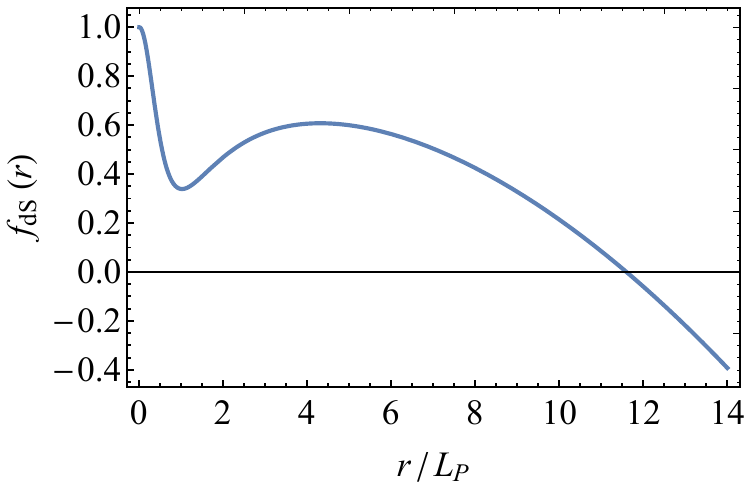}%
\caption{The ultracold metric potential in Planck units for $Q=e\,$, $\Lambda=0.02  \lpl^{-2}$ and $m\simeq 0.6 \mpl\,$.} 
\label{fig:fig12}
\end{figure}

The ultracold solution \cite{CJS98} is a super-degenerate black hole where all horizons coincide to a triple root $r_-=r_+=r_{\rm c}=\varrho_{\rm UC}\,$, as illustrated in Fig.~\ref{fig:fig12}.
The physical region of the positive-definite metric is $0<r\leq \varrho_{\rm UC}$ and the spacetime is like an inflating black hole. The ultracold system obeys the relation
\begin{equation} \label{ultra_rel}
f_{\rm dS}(\varrho_{\rm UC})=f'_{\rm dS}(\varrho_{\rm UC})=f''_{\rm dS}(\varrho_{\rm UC})=0 \,,
\end{equation}
from which we retrieve the black hole parameters ($m,Q,\Lambda$) with respect to the triple root. For the regularity of the line element, we  make the following coordinate transformation \cite{Sha13}:
\begin{equation} \label{ultra_trn}
\tau = \f{2\xi}{\epsilon f''_{\rm dS}(\varrho_{\rm UC})} \,, \qquad r=\varrho - \epsilon \cos \left( R \sqrt{\f{f_{\rm dS}''(\varrho_{\rm UC})}{2}}  \right)  ,
\end{equation}
where the coordinate $R$ has dimensions of length.
The metric potential then becomes 
\begin{equation} 
f_{\rm UC}(R) = \f{f''_{\rm dS}(\varrho_{\rm UC})}{2} \epsilon^2 \sin^2\left( R \sqrt{\f{f_{\rm dS}''(\varrho_{\rm UC})}{2}}  \right)
\end{equation}
and, for $\epsilon \rightarrow0\,$, the ultracold line element  reads
\begin{equation}
\d s^2 = \left( \f{\sin\left( R \sqrt{f_{\rm dS}''(\varrho_{\rm UC})/2}  \right)}{\sqrt{f_{\rm dS}''(\varrho_{\rm UC})/2}} \right)^2 \d \xi^2 + \d R^2 +  \varrho_{\rm UC}^2 \d \Omega^2 \,.
\end{equation}
In this case, the limit $f_{\rm dS}''(\varrho_{\rm UC}) \rightarrow 0$ 
simplifies the ultracold instanton as
\begin{equation}
\d s^2 =  \d R^2 + R^2 \d \xi^2  +  \varrho_{\rm UC}^2 \d \Omega^2 \,,
\end{equation}
which is the direct product of a flat $  R^2$ and a round 2-sphere $ \rm S^2$ of radius $\varrho_{\rm UC}\,$, namely $ R^2 \times \mathrm{S}^2$.

\section{Cosmological quantum production}
\label{sec:pbs}

So far, we have presented a geometric and thermodynamic overview of charged regular black holes, whether they exist in an asymptotically flat spacetime or within a cosmological background. In this final section, we will employ the no-boundary proposal \cite{HaH83,HHH19} to study the quantum mechanichal decay of empty de Sitter space into a black hole universe by constructing wavefunctions that describe both spacetimes. This proposal originates from the semi-classical theory of Euclidean quantum gravity \cite{GiH93}. In this theory,  Gibbons and Hawking attempted to approach semi-classically the quantum nature of gravity by combining Feynmann's path integral formalism with the classical gravitational action of a manifold. This approach allows us to describe  spacetime through a probability amplitude over all possible metric configurations $g_{\mu\nu}$ as
\begin{equation} \label{wvf}
\Psi = \int \mathcal{D}[g_{\mu\nu}] e^{i \mathcal{S}} \,,
\end{equation}
where $\mathcal{S}$ is the Lorentzian gravitational action. The above path integral is an oscillating, non-convergent integral and thus  cannot be solved. For this reason, one performs the technique of Wick-rotating the real time axis to an imaginary periodic time $\tau$, known also as Euclidean time, through the relation $t=-i\tau$. the periodicity  $\beta$ of the imaginary time is given by the inverse of the Hawking temperature ($\beta = T^{-1}$). The mathematical significance of this replacement is that the Lorentzian action  turns into a Euclidean action $I$ via the relation $\mathcal{S}=i I $, thereby replacing the oscillating behaviour of \eqref{wvf} with a convergent integral. 

In addition to the above,  the concept of a Euclidean manifold requires a  \textit{false vacuum}, for instance the Higgs vacuum \cite{BRM15}, which is responsible for generating this manifold. Subsequently,  half of the Euclidean section matches half of the  Lorentzian part through an analytical continuation process. This  essentially represents the tunnelling process in gravity. To be more precise, the false vacuum is assigned a probability  to nucleate a pair of Euclidean universes due to quantum fluctuations and, at a certain instant in time, these manifolds analytically continue to their Lorentzian counterparts \cite{BoHb96}. The probability of this quantum production is given by $P=|\Psi|^2$.  
Nevertheless, even a Euclidean path integral can sometimes be challenging to solve. For this reason, we search for saddle points of the Euclidean action, known as gravitational instantons. An instanton   represents  the classical path of the action  and provides the dominant contribution to the solution. This enables us to approximate the path integral with the exponent of the instanton action. Therefore, in the semi-classical approximation, the wavefunction \eqref{wvf} becomes
\begin{equation}
\Psi \approx e^{-I} ,
\end{equation}
where $I$ denotes from now on the instanton action. In the spirit of the above picture, we can compare the probability $P_{\rm bg}$ of a source-free cosmological space with the probability $P_{\rm obj}$ of the object that is nucleated inside this space. In doing so, we construct a pair creation rate for gravitational objects that reads
\begin{equation} \label{rate}
\Gamma = \frac{P_{\rm obj}}{P_{\rm bg}} = \frac{e^{-2I_{\rm obj}}}{e^{-2I_{\rm bg}}} \,.
\end{equation}
The rate \eqref{rate} represents a probability rate, indicating whether the probability of an object, like a black hole, is either suppressed or favoured compared to the empty background. 
If $\Gamma < 1\,$, the rate is suppressed, suggesting that the background is more probable to exist. Conversely, if $\Gamma > 1\,$, the rate is unsuppressed, favouring the production of the object. When $\Gamma = 1\,$, equal probabilities are expected for both the empty universe and the universe containing the object. Let us note here that the probabilities mentioned in \eqref{rate} are not normalized and  should be divided by a normalization factor. However, this normalization factor cancels out, leaving the above expression of the rate unchanged. Since we have the possibility of creating two distinct manifolds, then the normalized probability of the object should be
\begin{equation}
\mathcal{P}_{\rm obj} = \frac{|\Psi_{\rm obj}|^2}{|\Psi_{\rm obj}|^2+|\Psi_{\rm bg}|^2} = \frac{\Gamma}{\Gamma+1}
\end{equation}
and for the background
\begin{equation}
\mathcal{P}_{\rm bg} = \frac{|\Psi_{\rm bg}|^2}{|\Psi_{\rm obj}|^2+|\Psi_{\rm bg}|^2} = \frac{1}{\Gamma+1} \,.
\end{equation}
In the case where the rate is much smaller than unity ($\Gamma \ll 1$), the normalized probability of the object approximates the value of the rate ($\mathcal{P}_{\rm obj} \approx \Gamma$).

\begin{figure}[ht!] 
\begin{center}
\includegraphics[width=0.5 \textwidth]{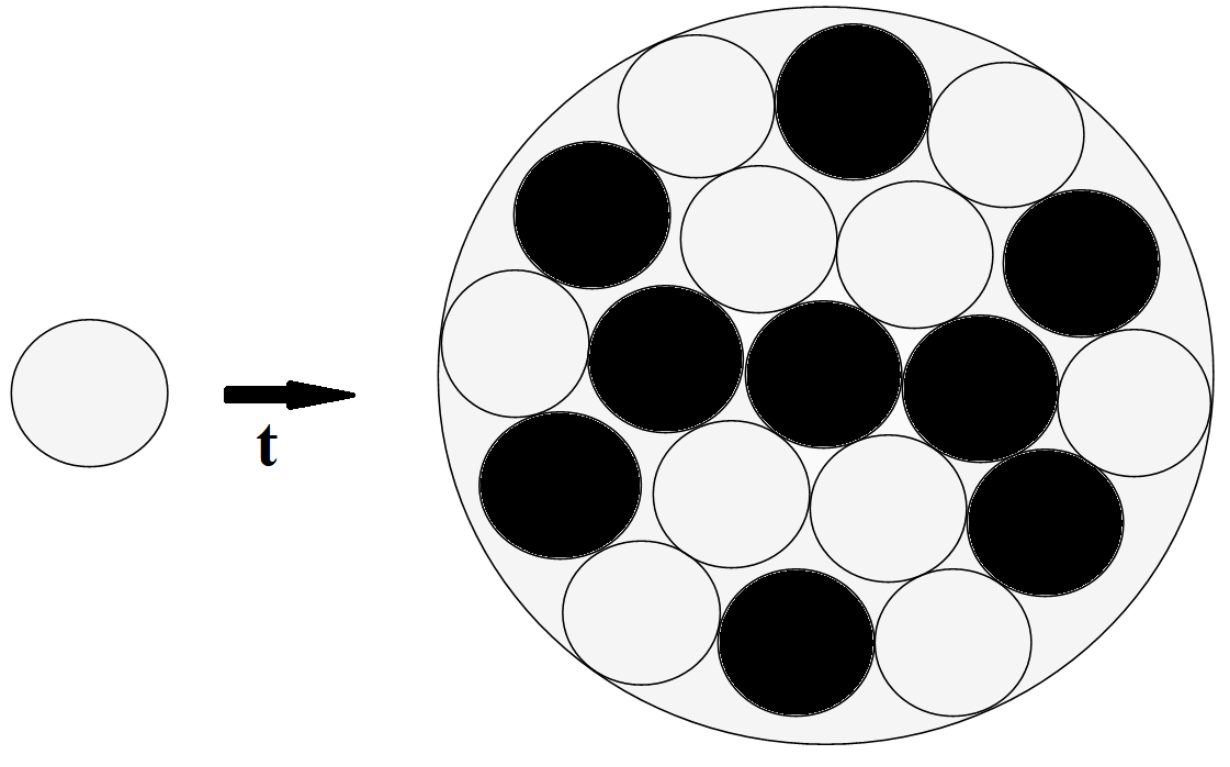}
\caption{ Schematic expansion of the universe during
one Hubble time and the bubbles (black) where black hole nucleation occurs.}
\label{fig:fig13}
\end{center}
\end{figure}

Bousso and Hawking investigated for the first time the quantum production of ordinary Schwarzschild-de Sitter  black holes back in the early universe \cite{BoH95,BoH96}. In their model they used the ideal solution of a Nariai black hole where the event horizon equals the size of the universe. The corresponding rate is exponentially suppressed for present times but it gives a substantial probability for  primordial black hole (PBH) population during inflation.  One may ask: \textit{how is it possible to talk about a population when the  metric refers to a single black hole?} The answer to this question lies in the concept of Linde's Eternal Chaotic Inflation \cite{Lin83,Gut00,Lin07},   according to the following description:

Let us assume that during the inflationary epoch, the universe had the size of a single Hubble volume unit. After a Hubble time the universe expanded and became $e^3\approx 20$ times bigger, increasing its Hubble volume to 20 units. According to de Sitter no hair theorem \cite{BoH96b}, inflationary expansion occurs independently in different regions of space because the inflaton field fluctuates and grows independently in each of these regions. However, Bousso and Hawking \cite{BoH96} have  also taken into account possible fluctuations that might affect the spacetime topology, leading to the formation of black holes in each of these regions. As a result, instead of considering one big Hubble  volume of 20 units, we consider twenty small Hubble volumes of 1 unit, with each having its own probability to nucleate a black hole, as depicted in Fig~\ref{fig:fig13}.
 In other words, the universe resembles a ``puzzle'' of sub-spaces (like bubbles), each containing, or not, a  black hole.
Being that said, the black hole metric refers to each of these sub-spaces independently. From this perspective, the pair creation rate $\Gamma$ can be interpreted as a particle rate, representing the number of black holes formed per Hubble volume during one Hubble time. The process of quantum black hole production seems to be a consequence of the violent expansion of the early universe. A helpful heuristic analogy is to imagine the rapid stretching of an elastic surface, like that of a balloon, which could result in the creation of holes on it.

Extensions of Hawking's idea have been made for charged \cite{MaR95}, rotating \cite{BoM99}, noncommutative \cite{MaN11}, higher dimensional \cite{DiL04} and $f(R)$ black holes \cite{DNT18}, all leading to the conclusion of a substantial pre-inflationary  black hole population. However, it is anticipated that this population would have been washed out by the rapid exponential expansion of the early universe, leaving no observable repercussions, just like in the case of magnetic monopoles. On the other hand, lower-dimensional scenarios involving a dilaton coupling could potentially result in a post-inlationary population of lower-dimensional PBHs \cite{DNT20}. 

Recently \cite{CNT25}, it has been shown that for any ultraviolet improved and neutral black hole metric, the rate is not so vastly suppressed even after inflation, resulting in a significant number of Planckian black holes in the present universe. This number is estimated to be on the order of $10^{60}$, contributing to a total mass that is close to the dark matter content. Consequently, the existence of nucleated, uncharged Planckian black holes could potentially offer an explanation for dark matter. This scenario requires extending the aforementioned pre-inflationary conjecture up to the present epoch. In other words, it is conjectured that as the universe continues to expand after inflation with the residual value of $\Lambda$, quantum fluctuations may still allow for the formation of black holes in different regions. These regions have sizes comparable to that of the ``baby universe" as it existed at the end of inflation, when the inflaton field relaxed to a minimum potential value that determined the size of one ``bubble" \cite{HaH17}. Based on these results, we follow the same procedure as in \cite{CNT25} but extend our investigation to explore the production of non-singular charged black holes inspired by NED.

 We start the current analysis with the Euclidean version of \eqref{action} along with the proper boundary term:
\begin{equation} \label{Euc_action}
I =  - \int \limits_{\mathcal{M}} \d ^4 x \sqrt{g} \left(  \frac{R-2\Lambda }{16 \pi \lpl ^2}   - \frac{\mathcal{L}}{4\pi}  +  \mathcal{L}_{\rm m} \right) - \frac{1}{4\pi } \int\limits_{\partial \mathcal{M}} \d x^3 \sqrt{h} \ \mathcal{L}_{F} F^{\mu\nu} \eta_{\mu}A_{\nu} \,,
\end{equation}
where $\eta_{\nu}$ is the  normal unit vector of the chosen boundary $\partial \mathcal{M}$ and $h$ is the determinant of the induced metric. Since we are working within a de Sitter background,   the cosmological horizon $r_{\rm c}$ naturally serves as an external boundary, interpreted as the zero-momentum initial data for the Lorentzian extension. This implies that the Gibbons-Hawking-York boundary term is  absent from the action due to the vanishing extrinsic curvature on the cosmological horizon. The only surviving boundary term is the second term appearing in \eqref{Euc_action} that fixes the charge on the boundary in a canonical ensemble. It is evident that, in comparison to the boundary term describing a classical charged black hole derived in \cite{HaR95}, the term obtained here contains an additional function, namely the derivative $\mathcal{L}_{F}\,$.

To proceed with the above integral calculation, we derive the form of the  Ricci scalar by tracing the Einstein field equation \eqref{field_eqs} and utilizing the relations \eqref{pressures}:
\begin{equation}
R= 4 \Lambda + 16\pi \lpl^2 \left( 2 \rho_{\rm m}(r) + \f{r}{2} \rho '_{\rm m}(r) + 2 \rho_{\rm e}(r) + \f{r}{2} \rho '_{\rm e}(r)\right) .
\end{equation}
To compute the matter Lagrangian, we will use the anisotropic fluid approach described in \cite{MaN11}. According to this approach, the on-shell matter Lagrangian, which obeys a stress-energy tensor as in \eqref{stress_tensors2}, is given by
\begin{equation}
\mathcal{L}_{\rm m} = 2 p_r^{(\rm m)}(r) - p_{\bot}^{(\rm m)}(r) = - \rho_{\rm m}(r) + \f{r}{2} \rho '_{\rm m}(r) .
\end{equation}
The NED Lagrangian is related to the tangential electric pressure. More precisely, by comparing the expressions \eqref{lagrang} and \eqref{p_e}, we obtain
\begin{equation}
\mathcal{L} = - 4\pi p_{\bot}^{(\rm e)}(r) = 4\pi \left(  \rho_{\rm e}(r) + \f{r}{2} \rho '_{\rm e}(r) \right) .
\end{equation}
By substituting the above three expressions into the parenthesis of the first integral in \eqref{Euc_action}, we find 
\begin{equation}
 \frac{R-2\Lambda}{16 \pi \lpl ^2}   - \frac{\mathcal{L}}{4\pi}  +  \mathcal{L}_{\rm m} = \f{\Lambda}{8\pi \lpl^2} + (r \rho_{\rm m}(r) )' + \rho_{\rm e}(r) \,,
\end{equation}
where the expressions for the matter and the electric energy density are given by \eqref{matter_d} and \eqref{p_e}, respectively. Moreover, integrating over the two angular coordinates and the imaginary time gives a factor of $2\pi \beta\,$ since the integration over $\tau$ is performed from 0 to $\beta/2$. This occurs because half of the instanton contributes to the solution, matching the other half of the Lorentzian section, as previously discussed in this section.

Switching our interest to the surface integral in \eqref{Euc_action}, we fix our timelike boundary on the cosmological  horizon $r_{\rm c}\,$, at least for the lukewarm and the cold case. Therefore, the induced metric is characterized by the following  line element 
\begin{equation}
\d s^2 = - f(r_{\rm c})\d t + r_{\rm c}^2 \d \Omega ^2 .
\end{equation}
The normal unit vector $\eta_{\mu}$ perpendicular to the boundary is given by
\begin{equation}
\eta_{\mu} = \f{\delta^{r}_{\mu}}{\sqrt{g^{rr}}} = (0,\f{1}{\sqrt{f(r_{\rm c})}},0,0) 
\end{equation}
 and is  spacelike. 
 From the $tr-$component of the field strength $F_{tr}=E(r)=-\p_{r}A_{t}\,$,  we can calculate the  scalar potential 
 \begin{equation}
 A_t = \f{Q}{8\lpl} \left[ \f{5\lpl r^3 + 3 \lpl^3 r}{(r^{2}+\lpl^2)^2} - \f{5\lpl r_+^3 + 3 \lpl^3 r_+}{(r_+^{2}+\lpl^2)^2} -3 \arctan\left( \f{r}{\lpl} \right) + 3 \arctan\left( \f{r_+}{\lpl} \right) \right] ,
 \end{equation}
after integrating from the horizon $r_+$ up to some distance $r$ for regularity reasons. Upon using the above relations along with \eqref{eom1}, the integration of the surface integral simply yields
\begin{equation}
- \frac{1}{4\pi } \int \d x^3 \sqrt{h} \ \mathcal{L}_{F} F^{\mu\nu} \eta_{\mu}A_{\nu} = - \f{\beta}{2} Q A_{t} \Big|_{r_{\rm c}} .
\end{equation}
Overall, the Euclidean action is found to be
\begin{equation} \label{bh_action}
I = - \f{\beta \Lambda}{12\lpl^2} (r_{\rm c}^3 - r_+^3) - 2\pi \beta \int\limits^{r_{\rm c}}_{r_+} \d r \ r^2 \left[ r \rho '_{\rm m}(r)+ \rho_{\rm m}(r) + \rho_{\rm e}(r) \right] - \f{\beta}{2} Q A_{t} \Big|_{r_{\rm c}} .
\end{equation}
Indeed, the full expression of the action is rather lengthy and mathematically difficult to handle. For this reason, numerical results will be presented for each nucleation channel separately. Before proceeding further, it is important to mention that the empty background ($m=Q=0$) corresponds to an empty de Sitter space  and yields an instanton action
\begin{equation} \label{ds_action}
I_{\rm dS} = - \frac{3\pi}{2  \Lambda \lpl^2} \,.
\end{equation}

\subsection{Lukewarm channel}

As stated in the previous section, lukewarm spacetime features a non-degenerate  event horizon being in thermal equilibrium with the cosmological surface. By solving the lukewarm conditions \eqref{lw_eq},  each quantity can be expressed with respect to the two horizons, according to \eqref{lw_rel}. However, the relations \eqref{lw_rel} are extremely long and complex,  requiring numerical methods for the calculation of the lukewarm rate. 

As discussed in \cite{CNT25}, for the conventional inflationary model, the black hole number lying inside the particle horizon just at the end of inflation, after accounting for the relevant reduction in number density, will be given by
\begin{equation} \label{number_inf}
N_{\rm end} \sim 10^{-102} \mathcal{P}_{\rm in} \,,
\end{equation}
while the  black hole number surviving today after the end of inflation should be
\begin{equation} \label{number_today}
N_{\rm today} \sim 10^{72} \mathcal{P}_{\rm fin} \,.
\end{equation}
These numbers result from dividing the Hubble volume of the universe into specific initial and final ``bubbles'' for each period, determined by the value of the cosmological constant before and after inflation. The value of $\Lambda$ just at the end of inflation should be  
\begin{equation}
\Lambda_{\rm fin} \sim 10^{-18} \lpl^{-2} \,,
\end{equation}
after assuming 60 e-foldings, while before inflation $\Lambda$ admits Planckian values, i.e.,
\begin{equation}
 \Lambda_{\rm in}\sim 1 \ \lpl^{-2}\,.
\end{equation}
The two normalized probabilities $\mathcal{P}_{\text{in}}$ and $\mathcal{P}_{\text{fin}}$ refer to nucleated black holes in each of the initial and final ``bubbles" respectively. These probabilities are approximately equal to the lukewarm rates, namely
\begin{equation} \label{NG}
N_{\rm inf} \approx 10^{-102} \Gamma_{\rm in} \qquad \mathrm{and} \qquad  N_{\rm today} \approx 10^{72} \Gamma_{\rm fin} \,,
\end{equation}
provided the rates are much smaller than unity. In addition, it follows from \eqref{number_inf} that for a surviving pre-inflationary PBH population, the corresponding rate $\Gamma_{\rm in}$ must significantly exceed unity in order to compensate for the exponential reduction in their number density. However, this does not appear to be the case, as it would imply a catastrophic instability for de Sitter space. For this reason, we consider only black holes that remain viable within the present particle horizon from the period starting at the end of inflation. The coefficient in \eqref{number_today} suggests a rate on the order of $10^{-12}$ to account for the dark matter content through charged  black holes.

The value $\Lambda_{\rm fin}$ imposes strict limits on the size of the two horizons, $r_+=a$ and $r_{\rm c}=b$. From \eqref{lw_rel}, $\Lambda_{\rm fin}$ can only be achieved for Planckian values of $a$, which in turn implies an extremely large value for $b$ due to the causal structure of the lukewarm spacetime. We cannot achieve the corresponding value of $\Lambda_{\rm fin}$ if the event horizon is comparable to the cosmological horizon. The values of the two horizon surfaces are determined numerically to be
\begin{equation}
a \sim 1.0006 \lpl \qquad \mathrm{and} \qquad b \sim 10^{9} \lpl \,.
\end{equation}
These two values fix in turn the charge and the bare mass as 
\begin{equation}
Q^2 \sim \frac{1}{137} \qquad \mathrm{and} \qquad  m \sim 0.9176 \mpl \,.
\end{equation}
One must be cautious during the numerical calculation due to charge quantization, which results in  discrete values for the rate. It is evident though that the black hole cannot possess more than one elementary electric charge ($N=1$), in order to match the proper value of $\Lambda_{\text{fin}}$. By substituting the aforementioned expressions into the action \eqref{bh_action}, termed here as the lukewarm action $I_{\rm lw}\,$, and employing the expression
\begin{equation}
\Gamma_{\rm lw} = \exp\left[ -2(I_{\rm lw}-I_{\rm dS}) \right] , 
\end{equation}
 we numerically determine the post-inflationary lukewarm rate to be
\begin{equation}
\Gamma_{\rm lw}^{\rm (fin)} \sim e^{-10^{11}} .
\end{equation}
This rate indicates a vast suppression with no lukewarm black hole remnants today. Even if we attempt to work at the beginning of inflation or during the Planck era by using $\Lambda_{\rm in}$ and repeat the same numerical procedure, the lukewarm rate can reach at most the maximum value of 
\begin{equation}
\Gamma_{\rm lw}^{\rm (in)} \sim 10^{-35} ,
\end{equation}
which is not sufficient to produce a significant number of pre-inflationary PBHs that could survive to the present universe.

\subsection{Cold channel}
By solving the cold relation \eqref{cold_rel}, we can express the bare mass, the charge and the cosmological constant in terms of the cold radius $r_+=\varrho_{\rm C}$ and the cosmological horizon $r_{\rm c}=b\,$:
\begin{equation} \label{cold_exp}
m=m(\varrho_{\rm C},b) \,, \qquad Q^2=Q^2(\varrho_{\rm C},b) \,, \qquad \Lambda(\varrho_{\rm C},b) \,.
\end{equation}
These expressions are mathematically simpler compared to those in the lukewarm case, yet they remain  lengthy. Hence, we omit their analytical form here. In this scenario, the period is determined by the inverse of the Hawking temperature corresponding to the cosmological surface, as the spacetime is in thermal equilibrium with the radiation from the cosmological horizon. That means
\begin{equation}
\beta = T_{\rm c}^{-1} =\left(  \frac{\d f_{\rm dS}(r)}{\d r} \Big|_{r=b} \right) ^{-1} ,
\end{equation}
where $f_{\rm dS}(r)$ is given by \eqref{ds_metric}. After using the  expressions \eqref{cold_exp}, the final form of the period becomes solely dependent on $\varrho_{\rm C}$ and $b$. At the end of inflation, the value of $\Lambda$ is approximately fixed as before  at $\Lambda_{\rm fin} \sim 10^{-18} \lpl^{-2}\,$, thereby determining the following numerical approximate values for the two horizons:
\begin{equation}
 \varrho_{\rm C} \sim 1.0006 \lpl \qquad \mathrm{and} \qquad b \sim 10^{9} \lpl \,.
\end{equation}
Subsequently, the charge and mass of the cold black hole are numerically calculated to be
 \begin{equation}
Q^2 \sim \frac{1}{137} \qquad \mathrm{and} \qquad  m \sim 0.9176 \mpl \,.
\end{equation}
Calculating next the cold action
$I_{\rm C}\,$, we can derive the expression of the cold rate from
\begin{equation}
\Gamma_{\rm C} = \exp\left[ -2(I_{\rm C}-I_{\rm dS}) \right] .
\end{equation}
 The post-inflationary cold rate is numerically calculated  to be 
\begin{equation}
\Gamma_{\rm C}^{\rm (fin)} \sim e^{-10^{10}} .
\end{equation}
This rate is vastly suppressed, though to a lesser extent compared to the lukewarm rate. 
Consequently, no cold black hole production occurs after inflation. Even upon repeating the numerical analysis for Planckian values of 
$\Lambda$ before inflation, the highest rate achievable is approximately
\begin{equation}
\Gamma_{\rm C}^{\rm (in)} \sim 10^{-30} ,
\end{equation}
which is insufficient to generate a substantial pre-inflationary cold PBH population that would survive in the present visible universe.
 
\subsection{Nariai channel}

In the charged Nariai case, where  a degeneracy between the event  and the cosmological horizon exists, the calculation of the Nariai instanton action is  simplified by employing the coordinate transformation  \eqref{Nariai_trs}. The degeneracy occurs when $\epsilon = 0\,$. We recalculate anew the two integrals of the instanton action \eqref{Euc_action}, namely $I=I_{\mathcal{M}}+I_{\mathcal{\partial \mathcal{M}}}=I_{\rm N}\,$, while considering the mentioned coordinate transformation. For the first integral, we obtain 
\begin{equation}
 I_{\mathcal{M}} = \frac{2\pi \varrho ^2}{f''_{\rm dS}(\varrho) \lpl^2} \left[ \Lambda + \frac{Q^2 \lpl^2}{(\varrho ^2+\lpl^2)^2} + \frac{\lpl^2-8\varrho ^2}{2\varrho ^2 + \lpl^2} \left( \frac{1}{\varrho ^2} -\Lambda - \frac{Q^2 \lpl^2}{(\varrho^2+\lpl^2)^2} \right)  \right] ,
\end{equation}
after taking the limit $\epsilon \rightarrow 0$ and employing the  relations \eqref{nar_eq}.
Setting $Q=0$ and working in the classical regime,  the approximations $\varrho ^2 \approx 1/\Lambda$ and $f''_{\rm dS}(\varrho) \approx -2\Lambda$ hold true and so the above action coincides with the uncharged classical Nariai action  found by  Hawking and Bousso \cite{BoH95}, i.e., $I_{\mathcal{M}}\approx -\frac{\pi}{\Lambda \lpl^2}\,$. 

Regarding the second integral $I_{\mathcal{\partial \mathcal{M}}}$, we consider the surfaces $\xi=0$ and $\xi=\pi$ to be the boundary \cite{HaR95}. While the coordinate $\xi$ has a period of $2\pi$, choosing $\xi=\pi$ reflects the integration of the periodic time up to $\beta/2$. 
Integrating from the black hole horizon ($\chi = 0$) to the cosmological horizon ($\chi = \pi$) along
$\xi = 0\,$, and back along $\xi = \pi\,$, we find that the boundary action takes on the simple form of
\begin{equation}
I_{\mathcal{\partial \mathcal{M}}} = - \frac{4\pi Q^2 \varrho ^4}{f''_{\rm dS}(\varrho) (\varrho ^2+\lpl^2)^3}  \,.
\end{equation}
In the classical limit ($\varrho \gg \lpl$), the above boundary action coincides with that of Mann and Ross \cite{MaR95}. The  expression of the Nariai action  is then given by the sum $I_{\rm N}=I_{\mathcal{M}}+I_{\mathcal{\partial \mathcal{M}}}\,$.
An alternative expression of the Nariai action can be obtained by exploiting the relations \eqref{nar_m} and \eqref{nar_Q}, yielding $I_{\rm N}=I_{\rm N}(\varrho,\Lambda)$. However, this expression  is omitted here due to its  lengthy form.  The  Nariai rate is then given by
\begin{equation}
\Gamma_{\rm N} = \exp [-2I_{\rm N} + 2 I_{\rm dS}] \,.
\end{equation}
After inflation ($\Lambda=\Lambda_{\rm fin}$), the 
numerical value of the post-inflationary pair creation rate is roughly
\begin{equation}
\Gamma_{\rm N}^{(\rm fin)} \sim e^{-10^{18}} ,
\end{equation}
 leaving  no possibility for post-inflationary Nariai black hole production. Before inflation ($\Lambda=\Lambda_{\rm in}$), numerical analysis yields a rate on the order of
\begin{equation}
\Gamma_{\rm N}^{(\rm in)} \sim e^{-100} ,
\end{equation}
indicating the absence  of pre-inflationary Nariai PBHs. Overall, the Nariai channel is more suppressed compared to the lukewarm or cold channels.

\subsection{Ultracold channel}

The ultracold black hole represents a super degenerate manifold where all horizons coincide. Employing the relation \eqref{ultra_rel} and the transformation \eqref{ultra_trn}, we recalculate  the instanton action \eqref{Euc_action} anew. The manifold exhibits an internal boundary at infinity. For this reason, we will use $R=R_0$ in the calculations and then take the limit $R_0 \rightarrow \infty\,$. The boundary is defined from the surfaces $\xi=0$ and $\xi=\pi$ along with the semi-circle lying between them at distance $R=R_0\,$. Therefore, we integrate  from $R=0$ up to $R=R_0$ along $\xi=0$ and  back along $\xi=\pi\,$.
The first integral of \eqref{Euc_action} gives
\begin{equation}
I_{\mathcal{M}} = \frac{\pi R_0^2}{4 \lpl^2} - \frac{\pi Q^2 \varrho_{\rm UC} ^4 R_0^2}{(\varrho_{\rm UC} ^2+\lpl^2)^3}
\end{equation}
while the surface integral of \eqref{Euc_action} gives
\begin{equation}
I_{\partial\mathcal{M}} = \frac{\pi Q^2 \varrho_{\rm UC}^4 R_0^2}{2(\varrho_{\rm UC}^2+\lpl^2)^3} \,.
\end{equation}
In the classical regime ($ \varrho_{\rm UC} \gg \lpl$), the approximations $4Q^2 \Lambda \lpl^2 \approx 1$ and $\Lambda \varrho_{\rm UC}^2 \approx2 $ hold true and so the aforementioned two integrals coincide with those derived in \cite{MaR95} for the classical charged ultracold case. Then, the ultracold action reads
\begin{equation} \label{uc_action}
I_{\mathrm{UC}} = \frac{\pi R_0^2}{4\lpl^2} - \frac{\pi Q^2 \varrho_{\rm UC} ^4 R_0^2}{2(\varrho_{\rm UC} ^2+\lpl^2)^3} \,
\end{equation}
and the ultracold rate is given by
\begin{equation} \label{uc_rate}
\Gamma_{\rm UC} = \exp [-2I_{\rm UC} + 2 I_{\rm dS}] = \exp \left[  -\frac{\pi R_0^2}{2\lpl^2} +   \frac{\pi Q^2 \varrho_{\rm UC} ^4 R_0^2}{(\varrho_{\rm UC} ^2+\lpl^2)^3} - \frac{3\pi}{  \Lambda \lpl^2} \right]  \,.
\end{equation}
For large ultracold radius, the action \eqref{uc_action} tends to zero ($I_{\mathrm{UC}} \approx 0$), indicating a vast suppression in the production rate. The causal structure of the ultracold manifold implies that a decrease in $\Lambda$ is followed by an increase in the radius $\varrho_{\rm UC}\,$. Therefore, after inflation, all ultracold black holes have sizes much larger than the Planck length and consequently cannot be produced through the quantum decay of de Sitter space, as their creation rate is given by 
\begin{equation}
\Gamma_{\rm UC}^{(\rm fin)} \approx \exp\left[ - \frac{3\pi}{\Lambda_{\rm end} \lpl^2} \right]  \sim e^{-10^{18}} .
\end{equation}
One may also derive an expression for the ultracold action and the corresponding rate solely dependent on the radius $\varrho_{\rm UC}$ upon employing the  relations \eqref{ultra_rel}. However, we omit these expressions here due to their lengthy form. Before inflation, numerical analysis for Plankcian values of $\Lambda$ suggests the existence of  Planck-sized ultracold black holes each  possessing one unit of elementary charge. In other words, the charge is constrained to be approximately $Q^2 \sim 1/137$ for $\varrho_{\rm UC} \sim \lpl$, which once more results in a suppressed rate for pre-inflationary ultracold PBHs. Roughly speaking, by substituting $ \varrho_{\rm UC} \approx \lpl\,$, $\Lambda \approx \lpl^{-2}$ and $Q^2 \approx 1/137$ into the expression of the rate \eqref{uc_rate}, we conclude to a suppressed pre-inflationary rate
\begin{equation}
\Gamma_{\rm UC}^{(\rm in)} \approx \exp \left[ - \frac{135\pi R_0^2}{274\lpl^2}   - 3\pi  \right] .
\end{equation}
Upon letting $R_0 \rightarrow \infty\,$, the rate tends to zero ($\Gamma_{\rm UC}^{(\rm in)} \rightarrow 0$). 
 
\section{Summary and outlook}
\label{sec:concl}

In this work, we have integrated the concept of nonlinear electrodynamics  with the existence of the minimal cut-off length, which is anticipated to modify the behaviour of gravity and electromagnetism at Planckian scales compared to the classical scenario. The presence of a charged matter source experiencing both NED and non-local effects results in a novel black hole solution, free from gravitational and electric ultraviolet pathologies. At large distances, the black hole matches known classical configurations, whereas near the origin, the manifold approaches the structure of de Sitter space.  Moreover, the thermal Hawking radiation does not diverge at the center. Rather, it reaches a maximum near the Planck length, signalling a phase transition  from an unstable black hole to a stable remnant. The evaporating black hole eventually reaches an extremal  stable configuration with  zero temperature and entropy, a fact that is in agreement with the third law of thermodynamics.

As a next step, we investigated the phenomenology of the  regular charged AdS black hole. This cosmological configuration exhibits an asymptotic behaviour at infinity, allowing us to define a total electro-gravitational mass. An observer standing in front of the black hole is in thermal equilibrium with the environment, since the cosmological surface lies at spatial infinity and so its thermal contribution can be ignored. This allows us to extend the phase space by identifying the negative $\Lambda$ with the thermodynamic pressure acting on the black hole.  This identification leads to an equation of state for the black hole which appears similarities with that of a Van der Waals gas. In conclusion, the black hole undergoes a first order phase transition between a small and a large stable configuration, mimicking this way the liquid/gas transition of a real gas.

Last but not least, we examined the regular charged    black hole in de Sitter space. In this case, the horizon structure is richer than in the other two asymptotic cases, yielding one, two, or three apparent singularities. The three-horizon topology is known as the lukewarm black hole. The two-horizon topologies correspond to the degenerate cases, known as the cold black hole and the Nariai black hole. Meanwhile, the single horizon topology, known as the ultracold black hole, represents a super-degenerate case where all three horizons coincide. The degenerate manifolds are in thermal equilibrium. Similarly, thermal equilibrium can be established between distinct horizons (lukewarm black hole) if the surface gravities of the event and cosmological horizons are equal in magnitude but opposite in sign. This can be achieved through a proper relation between the black hole parameters. In the Euclidean section, these four de Sitter cases represent regular  spacetimes, known as gravitational instantons, that can be pair created from the decay of de Sitter space, in accordance with the no-boundary proposal. Numerical analysis shows that the pair creation rate is vastly suppressed in all cases, whether evaluated before or after inflation. As a result, no viable remnants survive in the present universe. This  contrasts with the pair production of  Planckian uncharged black holes, whose present-day abundance can account for the observed dark matter content \cite{CNT25}. The key difference lies in the charged case, where an additional boundary term in the Euclidean action leads to significantly stronger suppression, eliminating any observable consequences for Planckian charged black holes.

Concluding, a more realistic scenario would involve rotations and the impact of the minimal length or NED  on all three parameters: mass, charge and rotation. The cosmological rotating and charged solution would thus present an intriguing subject for thermodynamic analysis and the study of its cosmological quantum production. We leave this field open for future research.

\subsection*{Acknowledgments}

The author would like to thank Piero Nicolini for reviewing the manuscript and offering insightful comments.


\end{document}